\documentclass{article}
\usepackage{arxiv}
\usepackage[utf8]{inputenc}

\usepackage[british]{babel}
\usepackage{amsmath}
\usepackage{amssymb}
\usepackage{amsthm}
\usepackage{mismath}
\usepackage{dsfont}
\usepackage[round]{natbib}
\usepackage{hyperref}
\usepackage{todonotes}
\usepackage[onehalfspacing]{setspace}
\usepackage{caption}
\usepackage{subcaption}

\newtheorem{theorem}{Theorem}[section]
\newtheorem{defi}[theorem]{Definition}
\newtheorem{lemma}[theorem]{Lemma}
\newtheorem{cor}[theorem]{Corollary}

\DeclareMathOperator{\clr}{clr}
\DeclareMathOperator{\argmax}{argmax}

\title{Principal component analysis in Bayes spaces for sparsely sampled density functions}

\author{Lisa Steyer \& Sonja Greven}
\date{\today}

\begin{document}
\maketitle

\begin{abstract}
This paper presents a novel approach to functional principal component analysis (FPCA) in Bayes spaces in the setting where densities are the object of analysis, but only few individual samples from each density are observed. We use the observed data directly to account for all sources of uncertainty, instead of relying on prior estimation of the underlying densities in a two-step approach, which can be inaccurate if small or heterogeneous numbers of samples per density are available. To account for the constrained nature of densities, we base our approach on Bayes spaces, which extend the Aitchison geometry for compositional data to density functions. For modeling, we exploit the isometric isomorphism between the Bayes space and the $\mathbb{L}^2$ subspace $\mathbb{L}_0^2$ with integration-to-zero constraint through the centered log-ratio transformation. As only discrete draws from each density are observed, we treat the underlying functional densities as latent variables within a maximum likelihood framework and employ a Monte Carlo Expectation Maximization (MCEM) algorithm for model estimation. Resulting estimates are useful for exploratory analyses of density data, for dimension reduction in subsequent analyses, as well as for improved preprocessing of sparsely sampled density data compared to existing methods. The proposed method is applied to analyze the distribution of maximum daily temperatures in Berlin during the summer months for the last 70 years, as well as the distribution of rental prices in the districts of Munich.
\end{abstract}
\newpage
%\tableofcontents

\section{Introduction}
A classic task in statistics is to estimate the underlying density from sample data, since density functions can be used to describe the distribution of real-valued random variables. However, when not all observations are identically distributed, but constitute repeated draws from a set of density functions $f_1, \dots, f_n$, e.g.\ for $n$ different individuals, distributional properties of these density functions themselves may be the actual target of a statistical analysis. Examples in which densities or distributions are considered the observational units exist in many different fields. These include the size distributions of different zooplankton in oceanology \citep{nerini}, the distributions of firm size in econometrics \citep{huynh}, mortality densities at different locations in epidemiology \citep{scimone}, and the densities of glucose levels among several diabetes patients in medical research \citep{matabuena}.

Probability density functions can be seen as a special case of functional data, but considering them as the unit of observation poses two major challenges. First, any density function must be non-negative and integrate to one to be valid. Second, in practice, density functions are often unobserved and accessible only through discrete samples. That is, for each density function $f_i$, $i = 1, \dots, n$ there is usually only an independent and identically distributed sample $x_{ij} \sim f_i$, $j = 1, \dots, m_i$ available. Our goal is to develop a Principal Component Analysis (PCA) for densities that can handle both of these challenges. A PCA for density data is of interest for several reasons. First, resulting estimates are useful for exploratory analyses to better understand the main modes of variation in density data. Second, the resulting  dimension reduction allows to succinctly describe differences and trends in densities and the corresponding principal components (PCs) can be used as a parsimonious data-driven basis  in subsequent analyses, as common in functional data analysis \citep{yao, chiou, scheipl}. Third, the reconstructed densities resulting from the PCA can also be used as improved preprocessing of sparsely sampled density data compared to existing methods, if subsequent analysis methods need observed or reconstructed density data as input \citep{ scimone, maier}.

Existing research has primarily addressed one of the two challenges associated with studying densities as functions, while overlooking the other. Initially, \cite{kneip} used functional principal component analysis (FPCA) in the unbounded space $\mathbb{L}^2$ of quadratic integrable functions without considering the density constraints. They did, however, account for discretely observed data by estimating the covariance surface based on the combined observations from all densities. In recent years, researchers have directed their attention towards incorporating the intrinsic geometric constraints in the space of density functions. The following discussion provides an overview of their work. However, when dealing with discretely observed data, their primary approach has been to estimate the observed densities through preprocessing steps, such as aggregating the data using histograms or kernel density estimates, and ignoring the reconstruction uncertainty in the further analysis. 

To address the density constraints, several metrics have been considered for the space of probability density functions. In particular, the Wasserstein distance \citep{panaretos} for probability measures is widely used, while the Fisher-Rao metric \citep{srivastava} imposes a manifold structure on the space of density functions. Although statistical analysis can be performed directly on manifolds (e.g. geodesic PCA in Wasserstein space \citep{bigot}), it is often more convenient to map the densities to a space with a simpler structure, perform statistical analysis there, and then back-transform the results to the original density space. Various transformations have been considered, such as the log-hazard and log-quantile density transformations \citep{petersen} for mapping the density functions to a Hilbert space. In particular, \cite{hron} used the centered log-ratio (clr) transformation to obtain FPCA for densities. The clr transformation is particularly useful here because it defines a one-to-one mapping between the squared-log integrable (proper and improper) density functions and the separable Hilbert space $\mathbb{L}^2_0$, which represents the space of square-integrable functions that integrate to zero. This means that the clr transformation also induces a Hilbert space structure on the space of squared-log integrable (proper and improper) density functions, which is called Bayes Hilbert space \citep{egozcue,boogaart}.

This paper develops Functional Principal Component Analysis (FPCA) in the Bayes spaces for the setting where only samples from the set of densities of interest are available. Our approach utilizes the observed data directly for calculations instead of estimating the underlying densities beforehand in a two-step approach. Estimating density $f_i$ becomes particularly challenging when dealing with small sample sizes $m_i$ in each observation unit. To address this, a two-step approach is proposed by \cite{qiu} for cases with heterogeneous sample sizes. Firstly, the underlying process is estimated based on a subsample where each density is densely observed. Then, the remaining densities are estimated using the process estimated in the first step as a prior. However, this approach is only feasible when there exists a representative subset of densities that have been densely observed.

The advantage of our approach is its applicability even when all densities are sparsely observed. To achieve this, we incorporate the observed data of all densities in a maximum likelihood framework, treating the underlying densities as latent variables. To estimate the model, we utilize the Expectation-Maximization (EM) algorithm \citep{dempster}. As the expectation step in our framework is not analytically solvable, requiring a Monte Carlo approximation, we base the estimation of our model on the so-called Monte Carlo Expectation-Maximization (MCEM) adaptation of the EM algorithm \citep{wei}. 

We organize our contributions as follows. First, in Section \ref{sec:method}, we introduce the latent density model and develop an MCEM algorithm suitable for this scenario. Then, in Section \ref{sec:applications}, we apply our methodology in two different applications. First, in the context of climate change and changing (extreme) temperatures in particular, we study how the distribution of maximum temperatures during the summer months in Berlin has evolved over the last 70 years. Second, we look at the distribution of rental prices in the different districts of Munich. In Section \ref{sec:simulation}, we use a simulation to demonstrate that our method is particularly effective for analyzing densities when few observations drawn from them are available. Finally, we conclude the paper with a discussion in Section \ref{sec:discussion}.

\section{Principal component analysis (PCA) for densities based on individual samples} \label{sec:method}
We consider densities as elements of the Bayes Hilbert space, which has proven to be a valuable framework for modeling densities. In order to perform Principal Component Analysis (PCA) in the Bayes Hilbert space, which is a separable Hilbert space, we employ the Karhunen-Loève decomposition. While PCA was originally developed as a dimension reduction tool for finite-dimensional data, the Karhunen-Loève decomposition extends this concept to infinite-dimensional Hilbert spaces \citep{hsing_eubank}.

\subsection{PCA in Bayes Hilbert spaces}
We start with reviewing the structure of the Bayes Hilbert space. For simplicity and as it seems natural in most applications, we restrict ourselves in this work mainly to densities with respect to the Lebesgue measure $\lambda$ defined on a compact interval $I \subset \mathbb{R}$ although the construction can be done for general measures \citep{boogaart}. In Subsection \ref{subsec:compositional_data}, we briefly discuss how the case of compositional data can be treated in a similar way.
\begin{theorem}[Bayes Hilbert space \citep{egozcue}]
Let $B = \{f = \exp(g)| g \in \mathbb{L}^2(I) \}$ and consider the equivalence relation $f_1 \sim f_2 \Leftrightarrow \exists \alpha > 0: f_1 = \alpha f_2$ for $f_1, f_2 \in B$. Denote by $\mathcal{B} = B/_\sim$ the set of equivalence classes $[f]$ with $f \in B$. Then $\mathcal{B}$ equipped with the operations
\begin{itemize}
    \item[$\oplus$] (addition) given by the pertubation operator $[f_1] \oplus [f_2] = [f_1 \cdot f_2]$ for all $[f_1], [f_2] \in \mathcal{B}$,
    \item[$\odot$] (scalar multiplication) given by the powering operation $\alpha \odot [f] = [f^\alpha]$ for all $[f] \in \mathcal{B}, \alpha \in \mathbb{R}$ and 
    \item[$\langle \cdot, \cdot \rangle_\mathcal{B}$] the scalar product given via $\langle [f_1], [f_2] \rangle_\mathcal{B} = \frac{1}{2|I|} \int_{I} \int_{I} \log\left( \frac{f_1(x)}{f_1(y)} \right) \log\left( \frac{f_2(x)}{f_2(y)} \right) dx \ dy$ for all $[f_1], [f_2] \in \mathcal{B}$
\end{itemize}
is a separable Hilbert space.
\end{theorem}

As a separable Hilbert space the Bayes Hilbert space is isometrically isomorphic to any other infinite dimensional separable Hilbert space, in particular $\mathcal{B}$ is isometrically isomorphic to $\mathbb{L}^2_0$, the space of square-integrable functions integrating to zero, via the centered log-ratio transformation.
\begin{lemma}[Centered log-ratio transformation] \label{lem:clr_trafo}
The centered log-ratio (clr) transformation 
\begin{align} \label{eq:clr_trafo}
   \clr: \mathcal{B} \to \mathbb{L}^2_0, \quad [f] \mapsto \log(f) - \frac{1}{|I|} \int_I \log(f(x)) dx
\end{align}
is a bijective isometry with the inverse given via $\clr^{-1}(g) = [\exp(g)]$ for all $g \in \mathbb{L}_0^2$.
\end{lemma}

In particular, the clr transformation is well defined as $\clr([\alpha f]) = \clr([f])$ for all $f \in B$ and $\alpha \in \mathbb{R}$. A detailed proof for Lemma \ref{lem:clr_trafo} can be found in Appendix \ref{app:clr_trafo} (compare with \cite{boogaart} for densities on general measure spaces). This identification of the Bayes space $\mathcal{B}$ with $\mathbb{L}_0^2$ allows statistical modeling to be performed in $\mathbb{L}_0^2$ instead of directly in $\mathcal{B}$, allowing the application of techniques developed for functional data. In particular, \cite{hron} used this correspondence to introduce PCA in Bayes space, while \cite{scimone} and \cite{maier} exploit its use for regression purposes. 

Similarly, we will achieve a principal component decomposition of observed densities $f_1, \dots, f_n$, $n \in \mathbb{N}$ via considering them being the back-transforms of realizations $g_1, \dots, g_n$ of a stochastic process $\mathcal{G} = \{G(x)\}_{x \in I} \subset \mathbb{L}_0^2$ characterized by its mean function $\mu(x) = \mathbb{E}(G(x))$ and covariance kernel $K(x_1, x_2) = \text{Cov}(G(x_1), G(x_2))$ for all $x, x_1, x_2 \in I$. The Karhunen-Loève decomposition \citep{karhunen, loeve} then yields the functional principal component representation 
\begin{align} \label{eq:karhunen_loeve}
    G(x) = \mu(x) + \sum_{k = 1}^\infty Z_k\varphi_k(x)
\end{align}
where $\varphi_k$, $k \in \mathbb{N}$ are the orthonormal eigenfunctions of the covariance operator $\mathbb{L}_0^2 \to \mathbb{L}_0^2$, $g \mapsto \int_I K(x_1, \cdot) g(x_1) d x_1$ associated with the covariance kernel $K$ and uncorrelated principal component scores $Z_k$ of decreasing importance, with $\mathbb{E}(Z_k) = 0$ and $\text{Var}(Z_k) = \sigma_k^2$ the corresponding eigenvalues, $\sigma^2_1 \geq \sigma^2_2 \dots \geq 0$. For more details on this decomposition for second-order stochastic processes refer to \cite{hsing_eubank}.

For a given sample of (fully observed) functions $g_1, \dots, g_n$, $n \in \mathbb{N}$ the unknown parameters $\varphi_k, Z_k$ of this process could then be estimated via computing the eigendecomposition of the sample covariance operator associated with the sample covariance kernel $\hat{K}_n(x_1, x_2) = \frac{1}{n} \sum_{i = 1}^n (g_i(x_1) - \hat{\mu}(x_1))(g_i(x_2) - \hat{\mu}(x_2))$ with $\hat{\mu}(x) = \frac{1}{n} \sum_{i = 1}^n g_i(x)$ for all $x, x_1, x_2 \in I$. Thus, the eigenfunctions $\varphi_k$ could be estimated as the eigenfunctions $\hat{\varphi}_k$ of the sample covariance and the distribution of $Z_k$ as the empirical distribution of the factor loadings $z_{ik} = \int_I (g_i(x) - \hat{\mu}(x))\hat{\varphi}_k dx$, $i = 1, \dots, n$ for all $k = 1, \dots, N$ where $N$ is the number of non-zero eigenvalues of the sample covariance.

The correspondence of $\mathbb{L}_0^2$ and the Bayes Hilbert space via the clr transformation (Lemma \ref{lem:clr_trafo}) gives an analog principal component decomposition to Equation \eqref{eq:karhunen_loeve} for densities. The process $\{\clr^{-1}(G)(x)\}_{x \in I}$ is given as
\begin{align} \label{eq:karhunen_loeve_bayes}
    \clr^{-1}(G)(x) = \clr^{-1}(\mu)(x) \oplus \bigoplus_{k = 1}^\infty \exp(Z_k) \odot \clr^{-1}(\varphi_k)(x)
\end{align}
inheriting the properties of the decomposition of $\mathcal{G}$ in $\mathbb{L}_0^2$ via the clr transformation (Equation \eqref{eq:clr_trafo}). Namely, we obtain orthonormal eigenfunctions since $\langle \clr^{-1}(\varphi_k), \clr^{-1}(\varphi_l) \rangle_\mathcal{B} = \langle \varphi_k, \varphi_l \rangle_{\mathbb{L}_2} = 0$ for all $k \neq l$ and $\|\varphi\|_\mathcal{B} = \|\varphi\|_{\mathbb{L}_2} = 1$ for all $k \in \mathbb{N}$ and principal components scores $\exp(Z_k), k \in \mathbb{N}$ with $Z_k \perp Z_l$ and $\mathbb{E}(Z_k) = 0$ for all $k,l \in \mathbb{N}$. If we additional assume that $G$ is Gaussian, this also implies that the principal component scores $\exp(Z_k), k \in \mathbb{N}$ are independent (and therefore also uncorrelated) and we can compute their expectation as the evaluation of the moment generating function, i.e. $\mathbb{E}(\exp(Z_k)) = \exp(0.5\sigma_k^2)$, where $\sigma^2_k = \text{Var}(Z_k)$.

Note that elements in $\mathcal{B}$ are equivalence classes with respect to scalar multiplication. As a result, they either possess a unique representative which is a proper density function, or solely consist of improper densities. In practical applications, the focus is often on modeling proper density functions. Consequently, it becomes necessary to impose restrictions on $\mathbb{L}_0^2$ accordingly. Specifically, only those functions $g \in \mathbb{L}_0^2$ that satisfy $\int_I \exp(g(x)) dx < \infty$ are suitable for representing proper densities. This requirement can be effectively met, for example, by using spline representations for elements of $\mathbb{L}_0^2$ \citep{machalova, maier} since the exponential transformation of a spline is bounded, guaranteeing integrability over $I$. Furthermore, if we consider a finite set of bounded density functions $f_1, \dots, f_n$, such as those obtained from kernel density estimates or histograms, and apply principal component decomposition in the Bayes space \eqref{eq:karhunen_loeve_bayes}, we are modeling only functions in the span of the data, which are also bounded. Consequently, in such cases, the eigenfunctions (and any linear combinations of them) are inherently proper densities.

Each equivalence class $[f]$ in the subset of $\mathcal{B}$ constructed such that it contains only proper densities can be uniquely identified with the element $\tilde{f} \in [f]$ that satisfies $\int_I \tilde{f}(x) dx = 1$. To simplify the notation, albeit with a slight misuse, we will refer to the equivalence class by this particular element, $\tilde{f} = [f]$ in the following discussion.
Hence, if $[\exp(g)]$ contains proper density functions, i.e. if $\int_I \exp(g(x)) dx$ is finite, we denote by
\begin{align} \label{eq:inverse_clr_trafo}
     \clr^{-1}(g) =  \frac{\exp(g)}{\int_I \exp(g(x)) dx} 
\end{align}
the back-transformed element in $\mathcal{B}$ under the inverse clr transformation.

\subsection{Likelihood formulation assuming latent densities}
We have seen in the previous section that the correspondence of $\mathbb{L}_0^2$ and the Bayes Hilbert space via the clr transformation provides a convenient approach for conducting PCA on fully observed densities, since it allows estimation via first transforming the density functions to the Hilbert space $\mathbb{L}^2$ using the clr transformation, and then performing the principal component decomposition in this well-known function space. %By transforming the density functions to the Hilbert space $\mathbb{L}^2$ using the clr transformation, we can perform the principal component decomposition in this well-known function space. 
This procedure for fully observed density functions was previously suggested by \cite{hron}. However, their approach motivates the so-called simplicial principal component analysis solely as a maximization problem. In other words, they seek to find the projections of the observed densities that maximize the variance along their directions. While this leads to the same decomposition for fully observed density functions when the covariance kernel is estimated using the sample covariance kernel, the stochastic process perspective becomes especially useful in the more common scenario we have in mind.

We focus on analyzing densities that are neither directly observable nor can be satisfactorily estimated from observed data as a preprocessing step. Instead, we assume that we have access to samples $x_{i1}, \dots, x_{im_i}$, where $m_i \in \mathbb{N}$, drawn from each probability distribution with density $f_i$, where $i = 1, \dots, n$. Our objective is then to conduct Maximum-Likelihood estimation for the parameters $\mu$ and $K$ of the underlying process $\mathcal{G}$ in $\mathbb{L}_0^2$ based on these samples. By estimating these parameters, we can compute the eigenvalues and eigenfunctions of the estimated covariance operator, which allows us to obtain the principal component decomposition which directly yields the principal component decomposition in the Bayes space via the inverse clr transformation. To accomplish this, we need to make a distributional assumption for $G$. For simplicity, we assume that $G$ follows a Gaussian process with a finite Karhunen-Loève decomposition. More precisely, we assume the following model.

\begin{defi}[Latent density model] \label{defi:latent_density_model}
Let $GP(\mu, K)$ be a Gaussian process with mean function $\mu$ and covariance kernel $K$ taking values in a finite dimensional subspace $\mathcal{H} \subset \mathbb{L}_0^2$ consisting of bounded functions. Then we assume the following data generating process
\begin{align*}
    X_{ij} \overset{i.i.d.}{\sim} \clr^{-1}(G_i) = \frac{\exp(G_i)}{\int_I \exp(G_i(x)) dx} 
\end{align*}
with $G_i$ being independent replicates of $GP(\mu, K)$ for all $i = 1, \dots, n$, $n \in \mathbb{N}$, $j = 1, \dots, m_i$.
\end{defi}

Assuming a finite dimensional subspace $\mathcal{H}$ is not restrictive in practice, where observed data always lies in a finite dimensional subspace. This assumption allows us to apply maximum likelihood theory, since in this case the parameters $\mu$ and $K$ of the underlying Gaussian process can be described by a finite set of real-valued parameters and the likelihood has a similar form as the marginal likelihood of a mixed model. More precisely, we can formulate the following corollary (proof in Appendix \ref{app:finite_gauss_process}).

\begin{cor} \label{cor:finite_gauss_process}
For any orthonormal basis $e_1, \dots, e_N \subset \mathbb{L}^2$ with $\mathcal{H} \subseteq \text{span}\{e_1, \dots, e_N\}$  we have that $G \overset{i.i.d.}{\sim} GP(\mu, K)$ is equivalent to 
$G = \sum_{k = 1}^N \theta_k e_k$ with $\boldsymbol{\theta} = (\theta_1, \dots, \theta_N)^T \overset{i.i.d.}{\sim} \mathcal{N}(\boldsymbol{\nu}, \boldsymbol{\Sigma})$ for $\mu = \sum_{k = 1}^N \nu_k e_k$ and $K(x_1, x_2) = \sum_{k = 1}^N \sum_{l = 1}^N e_k(x_1) e_l(x_2) \Sigma_{kl}$, where $\boldsymbol{\Sigma} = \left(\Sigma_{kl} \right)_{k,l = 1, \dots, N}$ and $\boldsymbol{\nu} = (\nu_1, \dots, \nu_N)^T$.\\
If $\boldsymbol{v}_1, \dots, \boldsymbol{v}_N$ are the eigenvectors of $\boldsymbol{\Sigma}$ with corresponding eigenvalues $\sigma^2_1, \dots, \sigma^2_N$ then $\varphi_l = \sum_{k = 1}^N v_{lk} e_k$, $l = 1, \dots, N$ are the eigenfunctions of the covariance operator given by the covariance function $K$ with the same eigenvalues $\sigma^2_1, \dots, \sigma^2_N$, where ${\boldsymbol{v}}_l = ({v}_{l1}, \dots, {v}_{lN})$ for all $l = 1, \dots, N$. 
\end{cor}

With this equivalence, the latent density model \ref{defi:latent_density_model} can also be written as
\begin{align} \label{eq:latent_density_model}
    X_{ij} \overset{i.i.d.}{\sim} \clr^{-1}(G_i) = \frac{\exp(G_i)}{\int_I \exp(G_i(x)) dx}  \quad \text{with} \quad  G_i = \sum_{k = 1}^N \theta_{ik} e_k \text{ and } \boldsymbol{\theta}_i = (\theta_{i1}, \dots, \theta_{iN}) \overset{i.i.d.}{\sim} \mathcal{N}(\boldsymbol{\nu}, \boldsymbol{\Sigma})
\end{align} and estimation of the parameters $\mu$ and $K$ is equivalent to estimation of $\boldsymbol{\nu}$ and $\boldsymbol{\Sigma}$. Note that we do not assume span\{$e_1, \dots, e_N\} = \mathcal{H}$, we only need to cover $\mathcal{H}$, since for any finite basis $\{e_1, \dots, e_N\}$ in $\mathbb{L}^2$ the sum to zero constrain carries over to the parameters $\boldsymbol{\nu}$ and $\boldsymbol{\Sigma}$. That means if $\int_I e_k(x) dx = \int_I e_l(x) dx$ for all $k,l = 1, \dots, N$, one just needs to ensure that the entries in $\boldsymbol{\nu}$ as well as the entries of all eigenvectors $\boldsymbol{v}_1, \dots, \boldsymbol{v}_N$ of $\boldsymbol{\Sigma}$ sum to zero.

Corollary \ref{cor:finite_gauss_process} also implies that the maximum likelihood estimators will be asymptotically consistent, if  model \ref{defi:latent_density_model} is correctly specified, that is the image $\mathcal{H}$ of the the process $G$ is actually in $\text{span}\{e_1, \dots, e_N\}$. An interesting question for future research is whether in the case of misspecification, in particular when the image space $\mathcal{H}$ is not finite, a sequence of finite-dimensional subspaces of $\mathbb{L}^2$ can be constructed such that the corresponding estimators are asymptotically consistent in this case as well.

To estimate the finite dimensional parameters $\boldsymbol{\nu}$ and $\boldsymbol{\Sigma}$ of the latent density model via maximum likelihood, we need to maximise the likelihood function given the realizations  $\boldsymbol{x}_i = (x_{i1}, \dots, x_{im_i})^T$ from the random sample $\boldsymbol{X}_i = (X_{i1}, \dots, X_{im_i})^T$, $i = 1, \dots,n$. Let $G_i = \sum_{k = 1}^N \theta_{ik} e_k$ and $\boldsymbol{\theta}_i = (\theta_{i1}, \dots \theta_{iN})^T$ for all $i = 1, \dots, n$. Then the marginal likelihood for the parameters $\boldsymbol{\nu}$ and $\boldsymbol{\Sigma}$ is given as
\begin{align} \label{eq:likelihood}
   L(\boldsymbol{\nu}, \boldsymbol{\Sigma}| \boldsymbol{x}_1, \dots, \boldsymbol{x}_n) = \prod_{i = 1}^n  \int_{\mathbb{R}^N} \frac{ \exp( \sum_{j = 1}^{m_i} \sum_{k = 1}^N \theta_{ik} e_k(x_{ij})) p(\boldsymbol{\theta}_i| \boldsymbol{\nu}, \boldsymbol{\Sigma}) }{\left(\int_I \exp(\sum_{k = 1}^N \theta_{ik} e_k(x)) dx \right)^{m_i} } d \boldsymbol{\theta}_i.
\end{align}

For a detailed derivation, please refer to Appendix \ref{app:likelihood}. Maximizing this marginal likelihood can be seen as an empirical Bayes approach, where the prior for $\boldsymbol{\theta}_i$ is a multivariate normal distribution with mean  $\boldsymbol{\nu}$ and covariance $\boldsymbol{\Sigma}$. Note that by $p$ we denote a general density function, for example here $p(\boldsymbol{\theta}_i| \boldsymbol{\nu}, \boldsymbol{\Sigma})$ denotes the density of a multivariate normal distribution with parameters $\boldsymbol{\nu}$ and $\boldsymbol{\Sigma}$.

Due to the complicated nature of the likelihood function and the potential abundance of parameters in $\boldsymbol{\nu}$ and $\boldsymbol{\Sigma}$, numerical optimization of \eqref{eq:likelihood} is challenging. Therefore, in the following section, we use the Monte Carlo Expectation Maximization (MCEM) algorithm as a numerical method to effectively tackle this maximization problem.

\subsection{Model estimation using an MCEM algorithm}
The EM algorithm, developed by \cite{dempster}, addresses the challenge of maximum likelihood estimation in the presence of incomplete or missing data. This algorithm provides a framework for estimating the parameters of statistical models that involve unobserved or latent variables. It iteratively updates parameter estimates by incorporating both observed data and estimates of the missing data. In our specific context, we want to use the EM algorithm to handle latent, unobserved densities $f_1, \dots, f_n$, along with observed values $\boldsymbol{x}_1, \dots, \boldsymbol{x}_n$ sampled from these densities. Central to the EM algorithm is the notion of the expected complete-data log-likelihood, which in our case becomes
\begin{align} \label{eq:expected_log_like}
    Q(\boldsymbol{\nu}, \boldsymbol{\Sigma}| \boldsymbol{\nu}^{(h)}, \boldsymbol{\Sigma}^{(h)}) 
    &= \mathbb{E}(\log(p(\boldsymbol{x}_1, \dots, \boldsymbol{x}_n, \boldsymbol{\theta}_1 \dots, \boldsymbol{\theta}_n| \boldsymbol{\nu}, \boldsymbol{\Sigma}))) \nonumber \\
    &= \mathbb{E}(\log(p(\boldsymbol{x}_1, \dots, \boldsymbol{x}_n| \boldsymbol{\theta}_1 \dots, \boldsymbol{\theta}_n))) + \mathbb{E}(\log(p(\boldsymbol{\theta}_1 \dots, \boldsymbol{\theta}_n| \boldsymbol{\nu}, \boldsymbol{\Sigma}))) + \text{const.} \nonumber \\
    &= \sum_{i = 1}^n \mathbb{E}(\log(p(\boldsymbol{\theta}_i|\boldsymbol{\nu}, \boldsymbol{\Sigma}))) + \text{const.}
\end{align}
where the expectation is taken with respect to the conditional distribution $\boldsymbol{\theta}_i| \boldsymbol{x}_i,  \boldsymbol{\nu}^{(h)}, \boldsymbol{\Sigma}^{(h)}$ of the parameters $\boldsymbol{\theta}_i = (\theta_{i1}, \dots, \theta_{iN})^T$ of the latent densities for all $i = 1, \dots, n$ given the current estimates $ \boldsymbol{\nu}^{(h)}$ and $\boldsymbol{\Sigma}^{(h)}$ for $\boldsymbol{\nu}$ and $\boldsymbol{\Sigma}$. However, in the given setting, the conditional distribution needed to compute the expectation of the complete-data log-likelihood is not directly available, making the computation intractable. To address this challenge, \cite{wei} introduce the Monte Carlo Expectation Maximization (MCEM) algorithm, which uses a Monte Carlo approach to approximate the expected value in $Q$. Thus, for our particular use case, we need to generate samples of $\boldsymbol{\theta}_i| \boldsymbol{x}_i,  \boldsymbol{\nu}^{(h)}, \boldsymbol{\Sigma}^{(h)}$ for all $i = 1, \dots, n$. In the following, we outline the procedure for obtaining these samples and implementing the MCEM algorithm in our use case. Subsections \ref{estep} and \ref{mstep} detail the E- and the M-steps, respectively, \ref{subsubsec:initial_values} the selection of model space and initial values, and \ref{MCEM:algo} summarizes the complete algorithm.

\subsubsection{E-step} \label{estep}
For all $i = 1, \dots, n$ we approximate the conditional expectation $\mathbb{E}(\log(p(\boldsymbol{\theta}_i|\boldsymbol{\nu}, \boldsymbol{\Sigma})))$ where the expectation is taken with respect to $\boldsymbol{\theta}_i| \boldsymbol{x}_i,  \boldsymbol{\nu}^{(h)}, \boldsymbol{\Sigma}^{(h)}$ using importance sampling, which is a method for estimating properties of a target distribution by sampling from another, auxiliary distribution. In our case we sample for all $i = 1, \dots, n$, $r$ replicates of the parameters $\boldsymbol{\theta}_{i}$ of the latent densities from an auxiliary distribution with density $p_i^*(\boldsymbol{\theta}_i)$ instead of $\boldsymbol{\theta}_i| \boldsymbol{x}_i,  \boldsymbol{\nu}^{(h)}, \boldsymbol{\Sigma}^{(h)}$. This yields replicates $\boldsymbol{\theta}_{i}^{(1)}, \dots, \boldsymbol{\theta}_{i}^{(r)}$, $r \in \mathbb{N}$, and we approximate
\begin{align} \label{MCEM}
 \mathbb{E}(\log(p(\boldsymbol{\theta}_i|\boldsymbol{\nu}, \boldsymbol{\Sigma}))) \approx \sum_{t = 1}^r \frac{\omega_{it}}{\sum_{t = 1}^r \omega_{it}} \log(p(\boldsymbol{\theta}_{i}^{(t)}|\boldsymbol{\nu}, \boldsymbol{\Sigma}))
\end{align}
with weights $\omega_{it}, t = 1, \dots, r$ given as $\omega_{it} = \frac{p(\boldsymbol{\theta}_{i}^{(t)} | \boldsymbol{x}_i,  \boldsymbol{\nu}^{(h)}, \boldsymbol{\Sigma}^{(h)})}{p_i^*(\boldsymbol{\theta}_{i}^{(t)})}$ for all $i = 1, \dots, n$. For details on this method and a comprehensive treatment of several related Monte Carlo methods, see the book by \cite{hammersley}.

The key to this method lies in selecting an appropriate auxiliary distribution. To achieve this, we use the eigen decompostion of $\boldsymbol{\Sigma}^{(h)}$ with sorted eigenvalues ${\sigma_1^2}^{(h)} \geq \dots \geq {\sigma_N^2}^{(h)}$ and corresponding eigenvectors $\boldsymbol{v}_1^{(h)}, \dots, \boldsymbol{v}_N^{(h)}$. Then $\boldsymbol{\theta}_i \sim \mathcal{N}(\boldsymbol{\nu}^{(h)}, \boldsymbol{\Sigma}^{(h)})$ is equivalent to $\boldsymbol{z}_i = \boldsymbol{V}^{(h)}(\boldsymbol{\theta}_i - \boldsymbol{\nu}^{(h)}) \sim \mathcal{N}(\boldsymbol{0}, \text{diag}({\sigma_1^2}^{(h)}, \dots, {\sigma_N^2}^{(h)}))$, where $\boldsymbol{V}^{(h)} = (\boldsymbol{v}_1^{(h)}, \dots, \boldsymbol{v}_N^{(h)})$ is a matrix whose columns are the eigenvectors of $\boldsymbol{\Sigma}^{(h)}$. Thus sampling from $\boldsymbol{\theta}_i| \boldsymbol{x}_i,  \boldsymbol{\nu}^{(h)}, \boldsymbol{\Sigma}^{(h)}$ is equivalent to sampling from the conditional distribution of the scores $\boldsymbol{z}_i = (z_{i1}, \dots ,z_{iN})$ given as

\begin{align} \label{eq:z_cond_density}
    p(\mathbf{z}_i | \boldsymbol{x}_i, \boldsymbol{\nu}^{(h)}, \boldsymbol{\Sigma}^{(h)}) 
    &\propto  p(\mathbf{x}_i | \mathbf{z}_i, \boldsymbol{\nu}^{(h)}, \boldsymbol{\Sigma}^{(h)}) p(\mathbf{z}_i| \boldsymbol{\Sigma}^{(h)}) 
    =  p(\mathbf{x}_i | \boldsymbol{\theta}_i = \boldsymbol{V}^{(h)T} \mathbf{z}_i + \boldsymbol{\nu}^{(h)}) \prod_{k = 1}^N p(z_{ik}| {\sigma_k^2}^{(h)}) \nonumber \\
    &= \prod_{j = 1}^{m_i} \clr^{-1}\left(\sum_{k = 1}^N \nu_k^{(h)} e_k +  \boldsymbol{z}_i^T \boldsymbol{v}^{(h)}_k e_k\right)(x_{ij})\prod_{k = 1}^N p(z_{ik}| {\sigma_k^2}^{(h)}) \nonumber \\
    &= \frac{\exp\left( \sum_{j = 1}^{m_i} \left( \mu^{(h)}(x_{ij}) + \sum_{k = 1}^N \boldsymbol{z}_i^T \boldsymbol{v}^{(h)}_k  e_k(x_{ij}) \right) \right) }{\left( \int_I \exp\left( \mu^{(h)}(x) + \sum_{k = 1}^N \boldsymbol{z}_i^T \boldsymbol{v}^{(h)}_k  e_k(x)  \right) dx \right)^{m_i}} \prod_{k = 1}^N p(z_{ik}| {\sigma_k^2}^{(h)})
\end{align}

where $z_{ik}| {\sigma_k^2}^{(h)} \sim \mathcal{N}(0, {\sigma_k^2}^{(h)})$
for all $i = 1, \dots, n$ and $k = 1, \dots, N$. Here, $\mu^{(h)} = \sum_{k = 1}^N \nu_k^{(h)} e_k$ is the current estimate for the mean function and $g_i = \mu^{(h)} + \sum_{k = 1}^N \boldsymbol{z}_i^T \boldsymbol{v}^{(h)}_k  e_k$, $i = 1, \dots, n$ are the current predictions for the latent clr transformed densities. Here we again take the Bayesian perspective, where  $p(\mathbf{z}_i | \boldsymbol{x}_i, \boldsymbol{\nu}^{(h)}, \boldsymbol{\Sigma}^{(h)})$ is the posterior distribution for the prior $\mathcal{N}(\boldsymbol{\nu}^{(h)}, \boldsymbol{\Sigma}^{(h)})$. Note that this posterior is a proper distribution if the prior $\mathcal{N}(\boldsymbol{\nu}^{(h)}, \boldsymbol{\Sigma}^{(h)})$ is proper, that is if all eigenvalues of $\boldsymbol{\Sigma}^{(h)}$ are finite.
\begin{lemma} \label{lem:proper_density}
Let ${\sigma_1^2}^{(h)} < \infty$. Then $\int_{\mathbb{R^N}} p(\mathbf{z}_i | \boldsymbol{x}_i, \boldsymbol{\nu}^{(h)}, \boldsymbol{\Sigma}^{(h)}) d \mathbf{z}_i < \infty$ for all $i$.
\end{lemma}
A proof for this statement can be found in Appendix \ref{app:proper_density}. This also implies that $p(\mathbf{z}_i | \boldsymbol{x}_i, \boldsymbol{\nu}^{(h)}, \boldsymbol{\Sigma}^{(h)})$ is decreasing as $\|\mathbf{z}_i\| \to \infty$ and since it is also continuous, it attains its mode $\mathbf{z}_i^* \in \mathbb{R}^n$. This is not necessarily the case if $\mathcal{N}(\boldsymbol{\nu}^{(h)}, \boldsymbol{\Sigma}^{(h)})$ is improper (see Appendix \ref{app:counterex_no_mode} for a counterexample). 

Since we assume here that the prior distribution $\mathcal{N}(\boldsymbol{\nu}^{(h)}, \boldsymbol{\Sigma}^{(h)})$ is proper, the mode of the posterior distribution $\mathbf{z}_i^* = \argmax_{\mathbf{z}_i \in \mathbb{R}^N} p(\mathbf{z}_i | \boldsymbol{x}_i, \boldsymbol{\nu}^{(h)}, \boldsymbol{\Sigma}^{(h)})$ is attained. Hence we can choose a multivariate normal distribution centered around the mode as an auxiliary distribution for the scores $\mathbf{z}_i$. % = (z_1 , \dots, z_N)^T$. 
We further choose the variances to be proportional to the prior variances ${\sigma_1^2}^{(h)}, \dots, {\sigma_N^2}^{(h)}$. This means that for a tuning parameter $\lambda > 0$, we choose the auxiliary distribution $p_i^*(\boldsymbol{z}_i)$ to be $\mathcal{N}(\boldsymbol{z}_i^*, \lambda \text{diag}({\sigma_1^2}^{(h)}, \dots, {\sigma_N^2}^{(h)}))$. Usually we set $\lambda = 1$, but if one wants to explore the parameter space for $\mathbf{z}$ more or less, one can also set $\lambda$ larger or smaller. Consequently, once we compute the mode $\mathbf{z}_i^*$, sampling from the auxiliary distribution $p_i^*$ reduces to independently sampling each element of the vector $\mathbf{z}_i$ from a univariate normal distribution. To numerically compute the mode, i.e., the maximizer of \eqref{eq:z_cond_density} with respect to $\mathbf{z}_i$, it is useful to derive the gradient of its log transformation to apply a gradient descent algorithm.

\begin{lemma} \label{lem:grad}
The gradient of the log conditional density of the scores $\mathbb{R}^N \to \mathbb{R}$, $\mathbf{z}_i \mapsto \log(p(\mathbf{z}_i | \boldsymbol{x}_i, \boldsymbol{\nu}^{(h)}, \boldsymbol{\Sigma}^{(h)}))$ is given as
\begin{align*}
    \nabla  \log(p(\mathbf{z}_i | \boldsymbol{x}_i, \boldsymbol{\nu}^{(h)}, \boldsymbol{\Sigma}^{(h)})) &= \sum_{k = 1}^N \boldsymbol{v}^{(h)}_k \left( \sum_{j = 1}^{m_i} e_k(x_{ij}) - m_i \langle f_{\mathbf{z}_i}, e_k \rangle_{\mathbb{L}_2}\right) - \left(\frac{z_{il}}{{\sigma_l^2}^{(h)}} \right)_{l = 1, \dots, N} 
\end{align*}
where $f_{\mathbf{z}_i} = \clr^{-1}\left(\mu^{(h)} + \sum_{k = 1}^N \boldsymbol{z}_i^T \boldsymbol{v}^{(h)}_k  e_k  \right)$ for all $\mathbf{z}_i = (z_{i1}, \dots, z_{iN})^T \in \mathbb{R}^N$. 
\end{lemma}

A detailed derivation can be found in Appendix \ref{app:grad}. With this readily available gradient, finding the mode becomes numerically feasible and we can obtain i.i.d. samples for the scores $\mathbf{z}_{it}$ and corresponding weights $\omega_{it} \in \mathbb{N}$ for all $t = 1, \dots, r$ using the importance sampling described above.

The equivalence of conditionally sampling $\boldsymbol{\theta}_i$ or $\mathbf{z}_i$ also yields samples $\boldsymbol{\theta}_{i}^{(t)}$ from $\boldsymbol{\theta}_i| \boldsymbol{x}_i,  \boldsymbol{\nu}^{(h)}, \boldsymbol{\Sigma}^{(h)}$ for all $t = 1, \dots, r$ via $\boldsymbol{\theta}_{i}^{(t)} = \boldsymbol{\nu}^{(h)} + \boldsymbol{V}^{(h)T} \mathbf{z}_{it}$. Hence, we approximate the expected complete-data log-likelihood given in \eqref{eq:expected_log_like} by
\begin{align} \label{eq:Q_function}
    Q(\boldsymbol{\nu}, \boldsymbol{\Sigma}| \boldsymbol{\nu}^{(h)}, \boldsymbol{\Sigma}^{(h)})  \approx \sum_{i = 1}^n \sum_{t = 1}^r \frac{\omega_{it}}{\sum_{t = 1}^r \omega_{it}} \log(p(\boldsymbol{\theta}_{i}^{(t)}|\boldsymbol{\nu}, \boldsymbol{\Sigma})) + \text{const.}
\end{align}
using the Monte-Carlo approximation given in \eqref{MCEM} for the conditional expectation.

\subsubsection{M-step} \label{mstep}
The M-step updates the parameters $\boldsymbol{\nu}$ and $\boldsymbol{\Sigma}$ by maximizing this approximation \eqref{eq:Q_function} of the $Q$-function. The new estimates are then given by 
\begin{align*}
    \left(\boldsymbol{\nu}^{(h+1)}, \boldsymbol{\Sigma}^{(h+1)} \right) 
    = \argmax_{\boldsymbol{\nu}, \boldsymbol{\Sigma}} Q(\boldsymbol{\nu}, \boldsymbol{\Sigma}| \boldsymbol{\nu}^{(h)}, \boldsymbol{\Sigma}^{(h)}) \approx \argmax_{\boldsymbol{\nu}, \boldsymbol{\Sigma}} 
    \sum_{i = 1}^n \sum_{t = 1}^r \frac{\omega_{it}}{\sum_{t = 1}^r \omega_{it}} \log(p(\boldsymbol{\theta}_{i}^{(t)}|\boldsymbol{\nu}, \boldsymbol{\Sigma})).
\end{align*}
Since for the latent density model \eqref{eq:latent_density_model} we assume that $\boldsymbol{\theta}_{i}^{(t)}|\boldsymbol{\nu}, \boldsymbol{\Sigma}$ follows a multivariate normal distribution with mean $\boldsymbol{\nu}$ and covariance matrix $\boldsymbol{\Sigma}$, this optimization problem corresponds to a weighted maximum likelihood estimation of the mean and the covariance matrix. Remarkably, this maximization problem also arises when the EM algorithm is used to estimate a Gaussian mixture distribution, allowing us to derive the solution based on the computations performed for this problem \citep[e.g.][]{bishop}. Hence, we compute the updates for the parameters of our model as 
\begin{align*}
    \boldsymbol{\nu}^{(h+1)} 
    &= \frac{1}{\sum_{i = 1}^n \sum_{t = 1}^r \omega_{it}} \sum_{i = 1}^n \sum_{t = 1}^r \omega_{it} \boldsymbol{\theta}_{i}^{(t)} \\
    \boldsymbol{\Sigma}^{(h+1)} 
    &= \frac{1}{\sum_{i = 1}^n \sum_{t = 1}^r \omega_{it}} \sum_{i = 1}^n \sum_{t = 1}^r \omega_{it} (\boldsymbol{\theta}_{i}^{(t)} - \boldsymbol{\nu}^{(h+1)} )(\boldsymbol{\theta}_{i}^{(t)} - \boldsymbol{\nu}^{(h+1)} )^T.\\
\end{align*}
These are the weighted mean and weighted covariance matrix of the samples of the principal component scores $\boldsymbol{\theta}_{i}^{(t)}$, $i = 1, \dots, n$, $t = 1, \dots, r$.

\subsubsection{Selection of model space and initial values} \label{subsubsec:initial_values}
In order to apply our method to real-world problems, we first need to find a suitable model space as well as suitable initial values for the MCEM algorithm. We suggest using piecewise constant spline functions for modeling. Since the piecewise constant functions are dense in $\mathbb{L}^2$, they allow us to approximate any function in $\mathbb{L}_0^2$, and thus any density in $\mathcal{B}$, with arbitrary accuracy if the nodes are chosen to be on a fine grid.  Therefore, we fix a fine grid for the knots $\kappa_1, \dots, \kappa_{N+1}$ and choose as a basis the indicator functions which are one between two neighboring knots and zero elsewhere. That is $e_k = \mathds{1}_{[\kappa_k , \kappa_{k + 1}]}$ for all $k = 1, \dots, N$.

To obtain suitable initial values for $\boldsymbol{\nu}^{(0)}$ and $\boldsymbol{\Sigma}^{(0)}$, we propose to first estimate the latent densities $\hat{f}_1, \dots, \hat{f}_n$ by kernel density estimation. We then develop their clr transformations $\hat{g}_1, \dots, \hat{g}_n$ in our basis $e_1, \dots, e_N$. Subsequently, we estimate $\boldsymbol{\nu}^{(0)}$ and $\boldsymbol{\Sigma}^{(0)}$ as the empirical mean and covariance of the coefficients $\boldsymbol{\theta}_1 \dots, \boldsymbol{\theta}_n$ of $\hat{g}_1, \dots, \hat{g}_n$, respectively. This approach effectively restricts the model space to the span of the kernel density estimates. This could be a problem, for example, if only a small sample of densities is available, or if the kernel density estimates are close to zero in some parts of the support. 

In this case, an alternative would be to initially select a lower dimensional, smooth spline space for modeling. If we choose an orthonormal basis $e_1, \dots, e_N \in \mathbb{L}_0^2$, such as normalized versions of the orthogonal compositional splines suggested by \cite{machalova}, we can choose arbitrary values for the initial mean and covariance of the coefficients, for instance, $\boldsymbol{\nu}^{(0)} = \boldsymbol{0} \in \mathbb{R}^{N}$ and $\boldsymbol{\Sigma}^{(0)} = \mathbb{I}_N \in \mathbb{R}^{N \times N}$, the identity matrix. 

On the other hand, when the number of densities $n$ is large, which results also in a $N = n$ basis functions by the procedure given above, not only does computing the mode become a high-dimensional optimization problem, which is computationally demanding, also calculating the weights $\omega_{it}$ becomes unstable as in this case typically many of the variances ${\sigma_k^2}^{(h)}$ will be very small, thus the product $\prod_{k = 1}^N p(z_k| {\sigma_k^2}^{(h)})$ will be close to zero. In this case, we suggest reducing the dimensionality of $GP( \hat{\mu}^{(h)}, \hat{K}^{(h)})$ in each step $h$ by setting the variances ${\sigma_k^2}^{(h)} = 0$ for $k > N'$, where $N'$ is a chosen value such that $N' \ll N$. The specific value of $N'$ can be determined based on the desired proportion of variance explained, while the proportion of variance explained should be considerably greater than desired for the final PCA.

\subsubsection{Estimation of the density PCA using the MCEM algorithm} \label{MCEM:algo}
Subsections \ref{estep}-\ref{subsubsec:initial_values} derived all necessary ingredients to now use the MCEM algorithm to estimate the PCA. First, we initialize $\boldsymbol{\nu}^{(0)}$ and $\boldsymbol{\Sigma}^{(0)}$ according to \ref{subsubsec:initial_values}. Then we iterate the following two steps until the convergence criteria $\|\boldsymbol{\nu}^{(h+1)} - \boldsymbol{\nu}^{(h)} \|_{\mathbb{L}^2} < \epsilon$ and $\|\boldsymbol{\Sigma}^{(h+1)} - \boldsymbol{\Sigma}^{(h)} \|_{\mathbb{L}^2} < \epsilon$ for a threshold $\epsilon > 0$ are reached.

- E-step following \ref{estep},

- M-step following \ref{mstep}.

Estimates at convergence then approximate the maximum likelihood estimates of the observed data likelihood, that is $\hat{\boldsymbol{\nu}} \approx \boldsymbol{\nu}^{(h+1)}$ and  $\hat{\boldsymbol{\Sigma}} \approx \boldsymbol{\Sigma}^{(h+1)}$ . Finally, the equivalence given in Corollary \ref{cor:finite_gauss_process} yields estimates for $\mu$, $\varphi_k$ and $\sigma^2_k$ in \eqref{eq:karhunen_loeve} using the eigendecomposition of $\hat{\boldsymbol{\Sigma}}$, where $\hat{\boldsymbol{v}}_1, \dots, \hat{\boldsymbol{v}}_N$ are the eigenvectors of $\boldsymbol{\Sigma}$ with corresponding eigenvalues $\hat{\sigma}^2_1, \dots, \hat{\sigma}^2_N$. Then, the estimate for the mean is obtained as $\hat{\mu} = \sum_{k = 1}^N \hat{\nu}_k e_k$, where $\hat{\boldsymbol{\nu}} = (\hat{\nu}_1, \dots, \hat{\nu}_N)^T$. The estimates for the eigenfunctions are given as $\hat{\varphi}_l = \sum_{k = 1}^N \hat{v}_{lk} e_k$, where $\hat{\boldsymbol{v}}_l = (\hat{v}_{l1}, \dots, \hat{v}_{lN})$ for all $l = 1, \dots, N$ with corresponding eigenvalues $\hat{\sigma}^2_1, \dots, \hat{\sigma}^2_N$.

 The scores $Z_{ik}$, $k = 1, \dots, N$ for each density $f_i, i = 1, \dots, n$ are then predicted as the posterior mode given the estimates $\hat{\boldsymbol{\nu}}$ for the mean and $\hat{\boldsymbol{\Sigma}}$ for the covariance of the coefficients, that is $\hat{\mathbf{z}}_i = \argmax_{\mathbf{z}_i \in \mathbb{R}^N} p(\mathbf{z}_i | \boldsymbol{x}_i, \hat{\boldsymbol{\nu}}, \hat{\boldsymbol{\Sigma}})$ with $\hat{\mathbf{z}}_i = (\hat{z}_{i1}, \dots \hat{z}_{iN})$. This also yields predictions for the latent densities as  $\hat{f}_i = \clr^{-1}(\hat{g}_i)$ with $\hat{g}_i = \hat{\mu} + \sum_{k = 1}^N \hat{z}_{ik} \hat{\varphi}_k$ for all $i = 1, \dots, n$.

\subsection{PCA for compositional data as a special case} \label{subsec:compositional_data}
Although we have limited our considerations so far to densities with respect to the Lebesgue measure, it's important to recognize that our methodology can be seamlessly extended to densities with respect to arbitrary measures. In the following, we illustrate this via showing how our approach can be used to derive Principal Component Analysis (PCA) for compositional data, namely densities with respect to the discrete measure. Notably, our method has the advantage of being applicable to ``sparsely observed'' compositional data, known as count compositions in the compositional data literature \citep{filzmoser2018analyzing}, even when some of the categories have no observations  and without imputations.

The discrete measure on the power set $\mathcal{P}(\{A_1, \dots, A_k\})$ of a finite set of disjoint outcomes $A_1, \dots, A_k$ is given as $\eta = \sum_{k =1}^N \delta_{A_k}$, where $\delta_{A_k}(B) = 1$, if $B = A_k$ and $\delta_{A_k}(B) = 0$ else, for all $B \in \mathcal{P}(\{A_1, \dots, A_k\})$. Hence, every probability measure on $\mathcal{P}(\{A_1, \dots, A_k\})$ is given by a discrete density, i.e.\ probability mass function $f: \{A_1, \dots, A_k\} \to \mathbb{R}$ with respect to $\eta$ by the Radon-Nikodym Theorem. Here, the density $f$ is characterized solely by the values $\pi_k = f(A_k)$ for all $k = 1, \dots, N$, and since we consider only probability measures, it must hold that $\sum_{k = 1}^N \pi_k = 1$. That means $\mathcal{B}$, the set of densities with respect to the discrete measure on $\mathcal{P}(\{A_1, \dots, A_k\})$ can be identified with the simplex $\{ \boldsymbol{\pi} \in \mathbb{R}^N |  \sum_{k = 1}^N \pi_k = 1, \pi_k \geq 0 \ \forall k = 1,\dots, N\}$ and via the discrete centered log-ratio transformation $\boldsymbol{\rho} = \clr(\boldsymbol{\pi}) = (\log(\pi_1) - \frac{1}{N}\sum_{k = 1}^N \log(\pi_k), \dots, \log(\pi_N) - \frac{1}{N}\sum_{k = 1}^N \log(\pi_k))$ with the $N-1$ dimensional Hilbert space $\mathcal{H} = \mathbb{R}^N_0 = \{ \boldsymbol{\rho} \in \mathbb{R}^N | \sum_{k = 1}^N \rho_k = 0\}$ \citep{aitchison}. If we equip this space with the standard scalar product on $\mathbb{R}^N$, we obtain the Aitchison geometry on $\mathcal{B}$, which defines a Hilbert space structure for compositional data, i.e. for densities with respect to the discrete measure. 

In order to use our method to perform principal component analysis for observed count compositions $\boldsymbol{\pi_1}, \dots, \boldsymbol{\pi_n}$, we need to find a suitable basis representation of $\mathcal{H}$, as well as appropriate initial values for the mean and covariance of the corresponding basis coefficients. For densities with respect to the discrete measure, kernel density estimation cannot be employed due to the usual lack of an order relation and thus neighborhood structure to smooth over. We therefore rely on the alternative described in Subsubsection \ref{subsubsec:initial_values} and use an orthonormal basis of $\mathcal{H} = \mathbb{R}^N_0$. %to be able to choose arbitrary initial values for the mean and the covariance matrix. 

There are many options since % when it comes to choosing an orthonormal basis. This flexibility arises from the fact that 
an orthonormal basis can be obtained from any basis of $\mathbb{R}^N_0$ using Gram-Schmidt orthogonalization. For example, \cite{egozcue2003} suggest to use the following such basis %obtained by this process. They suggest to use
\begin{align*}
    e_k = \sqrt{\frac{k}{k + 1}}(\overbrace{k^{-1}, \dots, k^{-1}}^{k \text{ times}}, -1, 0, \dots, 0)^T
\end{align*}
with $k = 1, \dots, N$ as an orthonormal basis of $\mathbb{R}^N_0$, which yields an orthonormal basis of $\mathcal{B}$ via the inverse clr transform. With this basis choice we propose to set $\boldsymbol{\nu}^{(0)} = \boldsymbol{0} \in \mathbb{R}^{N - 1}$ and $\boldsymbol{\Sigma}^{(0)} = \mathbb{I}_N \in \mathbb{R}^{(N-1) \times (N-1)}$ as initial values for the mean and covariance of the basis coefficients and proceed with the MCEM algorithm as in the continuous case described above.  
\section{Applications} \label{sec:applications}
In this section, we demonstrate the applicability and advantages of our latent density-based PCA. We consider two different applications. In the first application, we analyze densities describing distributions of summer daily maximum temperatures in Berlin. In the second application, the latent densities describe the distribution of rental prices for each district in Munich.

\subsection{Distributions of daily maximum temperature in summer months per year}
According to the Copernicus Climate Change Service (C3S) report (\url{https://climate.copernicus.eu/esotc/2022/temperature}), 2022 was the second warmest year on record in Europe, with temperatures 0.9°C above the long-term average. In particular, the summer of that year set a new mark as the hottest on record, with temperatures 1.4°C above average, topping the previous warmest summer in 2021 by 0.3°C. 

These findings are consistent with the observed trend of increasing temperatures, an indicator of climate change. To refine the analysis of summer temperature trends, we focus not on average temperatures, but on trends in the entire distribution of daily maximum temperatures during the summer months of June, July, and August each year. That is, we consider as observational units densities that describe the distributions of daily maximum temperatures in summer. Thus, for every year, we treat the daily maximum temperature measured for the 92 days in June, July, and August as observations from these densities per year. The daily maximum temperature data  we use in this application have been collected from 1951 to 2022 at a single weather observatory, which is Berlin Tempelhof. Data and metadata are available at \url{https://www.ecad.eu}, provided by the ECA\&D project \citep{tank}. 

\begin{figure}[ht]
    \centering
    \includegraphics[width=\textwidth]{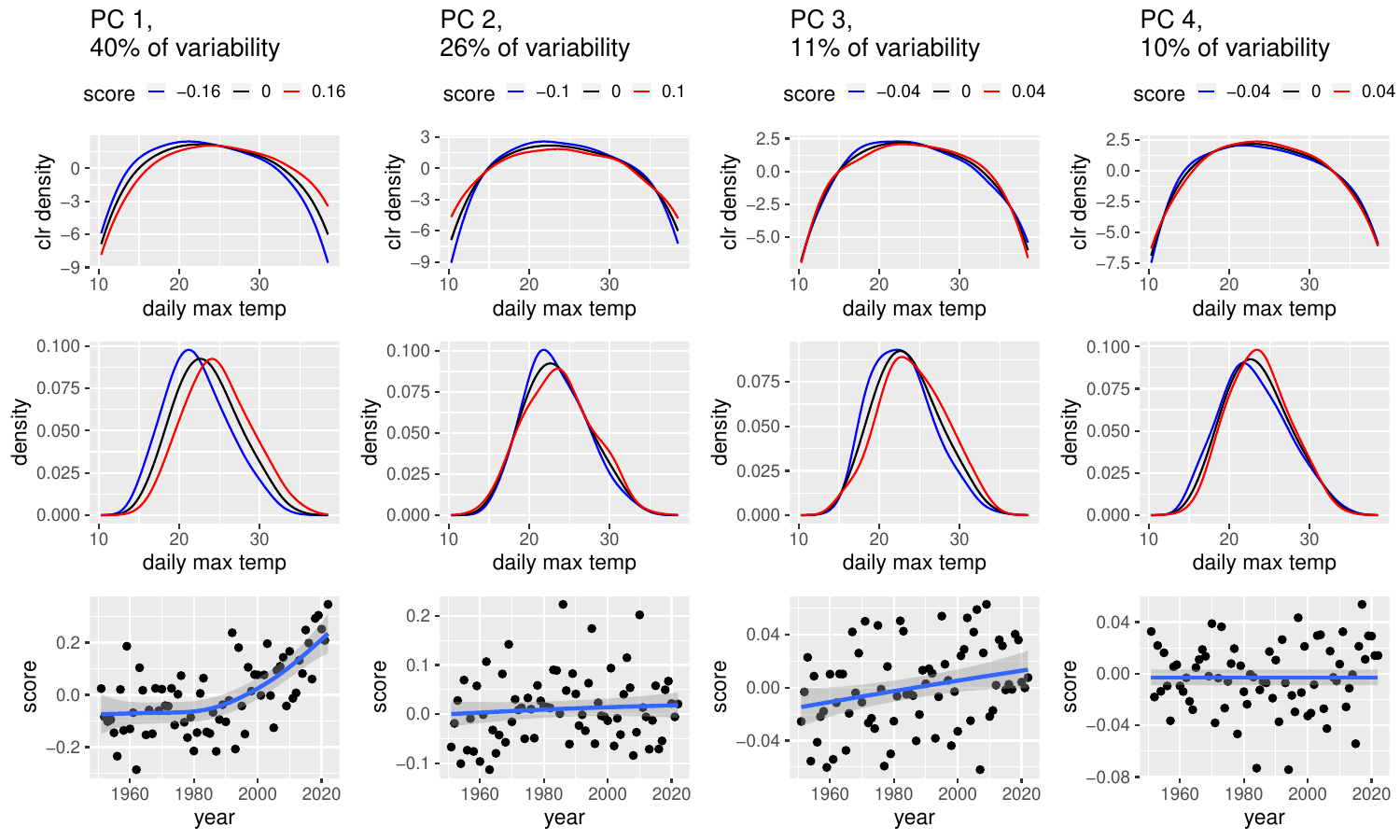}
    \caption{Latent density PCA for daily maximum temperature (°C). Top: Effect of adding/subtracting $\hat{\sigma}_k \hat{\varphi}_k$ to the clr transformed mean density $\mu$, where $\hat{\varphi}_k$ is the $k$th principal component, with corresponding eigenvalue $\hat{\sigma}_k^2$, $k = 1,2,3, 4$. Middle: Effect on the density level, i.e. $\clr^{-1}$ transformations of the functions in the top row. Bottom: Temporal trend of the corresponding predicted scores per year, with scatterplot smoother and pointwise confidence bands overlaid.}
    \label{fig:temperature_density_pca}
\end{figure}

To visualize and describe how the distribution of daily maximum temperature has changed over the period from 1951 to 2022, we proceed as follows. First, we obtain a low-dimensional representation of the latent daily maximum temperature densities using our latent density PCA and with kernel density estimates as initial estimates (Figure \ref{fig:temperature_densities} in the appendix) For technical details of the estimation please refer to Appendix \ref{app:temperature}). Then, in Figure \ref{fig:temperature_density_pca}, we visualize the first four principal components on clr level (top row), and transformed back to density level (middle row). In the bottom row, we plot the temporal trend of the corresponding predicted scores, overlaying a scatterplot smoother and pointwise confidence bands based on \cite{wood}.

This shows that adding a multiple of the first principal component, which explains 40\% of the total variability in latent densities, to the estimated mean mainly causes a rightward shift of the density. In particular, a positive value of the first principal component implies that high temperatures are more likely than in average years. Looking at the effect on clr level, we notice that this is especially true for temperatures above 35°C, which are more likely to occur in years with high first principal component scores. Notably, these first principal component scores show a clear increase over the time course from 1951 to 2022 (Figure \ref{fig:temperature_density_pca}, bottom left), meaning that hot and also very hot daily maximal temperatures in summer became more likely.

In contrast, the scores associated with the second and fourth principal components show no or almost no visible temporal trend. Adding these principal components to the estimated mean results in smaller changes in the shape of the density, with subtler shifts to the right in certain areas of the density. However, adding the third principal component to the estimated mean shifts the density towards experiencing moderate to hot temperatures (25°C-35°C) more frequently, while decreasing the likelihood of milder temperatures (15°C-25°C). The scores associated with this third principal component also show an increase over time.

\begin{figure}[ht]
    \centering
    \includegraphics[scale = 0.6]{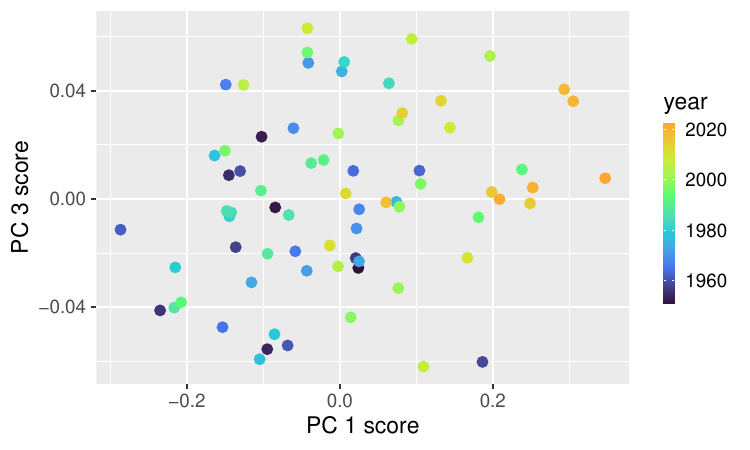}
    \caption{Scores associated with the first and third principal components for all years.}
    \label{fig:temperature_pc1_vs_pc3}
\end{figure}

This means that the trends of both the first and third principal component scores indicate that hot and very hot days are becoming more likely. When we plot both scores together (Figure \ref{fig:temperature_pc1_vs_pc3}) we see that all early years, corresponding to (dark) blue points, tend to lie in the bottom left corner, while recent years (yellow and orange points) are predominantly in the top right corner. Thus, recent years have high first principal component scores and/or high third principal component scores, which means the likelihood for hot and/or very hot days has been higher in these years than in earlier years. 

This application shows that the estimation of the latent density model \eqref{eq:latent_density_model} is suitable for identifying a small number of principal directions of variation in the data for densities when only discrete observations of each density are available. These principal components can then be used to visualize the data and/or for further analysis, such as to relate them to other scalar variables (in our case, the year of observation). 

However, in this homogeneous and only mildly sparse setting with 92 observations per density, differences to the simpler two-step approach with pre-smoothing (our starting values), a PCA obtained from the clr transformations of the kernel density estimates (see Figure \ref{fig:temperature_pca_pre_smooth}), are relatively small. One notable difference, though, is that in the latent density model, the trend of the score associated with the first principal component changes more rapidly around 1990, with a steeper increase since that time, than that of the first principal component scores obtained in the two-step approach. The trend in the third PC score is somewhat more pronounced as well, which may be due to unstable kernel density estimates in areas where the densities are close to zero. In contrast to this first application, our second application has more heterogeneous sample sizes and we will  observe how this unbalanced setting affects the estimation.

\subsection{Distributions of rental prices in the districts of Munich}
We consider data that was collected in Munich in 2019 (the most recent year) to construct an official rent index by estimating the rent given certain covariates of the apartments in a regression model. For details on the data collection, the available variables, estimation and interpretation of the rent index, see \cite[in German]{windmann}. A sub-dataset, which contains data on all 3255 apartments but only a subset of the covariates used for the official Munich rent index, is provided in the supplemental material of \cite{fahrmeir} and is also used here.

The goal of the analysis here is to describe the differences in the distributions of rental prices across the districts of Munich. To do this, we model the distribution of net rents per square meter, assuming a latent density for each district. Figure \ref{fig:rent_histograms} in Appendix \ref{app:rent} displays the histogram estimates and kernel density estimates (Gaussian kernel, bandwidth = 2) for the densities in each district. Table \ref{tab:munich_districts} shows that the number of observations $m_i$, $i = 1, \dots, 25$ per district varies considerably in this dataset. It ranges from 29 observations in district 23-Allach-Untermenzing to 261 observations in district 9-Neuhausen-Nymphenburg. Below, we will compare the estimation of our latent density model \eqref{eq:latent_density_model} for this heterogeneous sampling scheme to a two-step approach, where the densities are first estimated and then PCA is performed. 

\begin{table}[ht]
\begin{tabular}{rp{5cm}r}
  \hline
district $i$ & name & $m_i$ \\ 
  \hline
1 & Altstadt-Lehel & 79 \\ 
  2 & Ludwigsvorstadt-Isarvorstadt & 217 \\ 
  3 & Maxvorstadt & 219 \\ 
  4 & Schwabing-West & 241 \\ 
  5 & Au-Haidhausen & 230 \\ 
  6 & Sendling & 136 \\ 
  7 & Sendling-Westpark & 95 \\ 
  8 & Schwanthalerhöhe & 110 \\ 
  9 & Neuhausen-Nymphenburg & 261 \\ 
  10 & Moosach & 81 \\ 
  11 & Milbertshofen-Am Hart & 101 \\ 
  12 & Schwabing-Freimann & 198 \\ 
  13 & Bogenhausen & 165 \\ 
  \hline
\end{tabular}
\begin{tabular}{rp{5cm}r}
  \hline
district $i$ & name & $m_i$ \\ 
  \hline
  14 & Berg am Laim & 75 \\ 
  15 & Trudering-Riem & 92 \\ 
  16 & Ramersdorf-Perlach & 111 \\ 
  17 & Obergiesing-Fasangarten & 125 \\ 
  18 & Untergiesing-Harlaching & 151 \\ 
  19 & Thalkirchen-Obersendling-Forstenried-Fürstenried-Solln & 160 \\ 
  20 & Hadern & 64 \\ 
  21 & Pasing-Obermenzing & 107 \\ 
  22 & Aubing-Lochhausen-Langwied & 35 \\ 
  23 & Allach-Untermenzing & 29 \\ 
  24 & Feldmoching-Hasenbergl & 38 \\ 
  25 & Laim & 135 \\ 
   \hline
\end{tabular}
\caption{Munich districts: number $i$, name and number of observations per distric $m_i$\label{tab:munich_districts}}
\end{table}

\begin{figure}
\vspace{-0.3cm}
\centering
\begin{subfigure}[c]{\textwidth}
    \includegraphics[width=\textwidth]{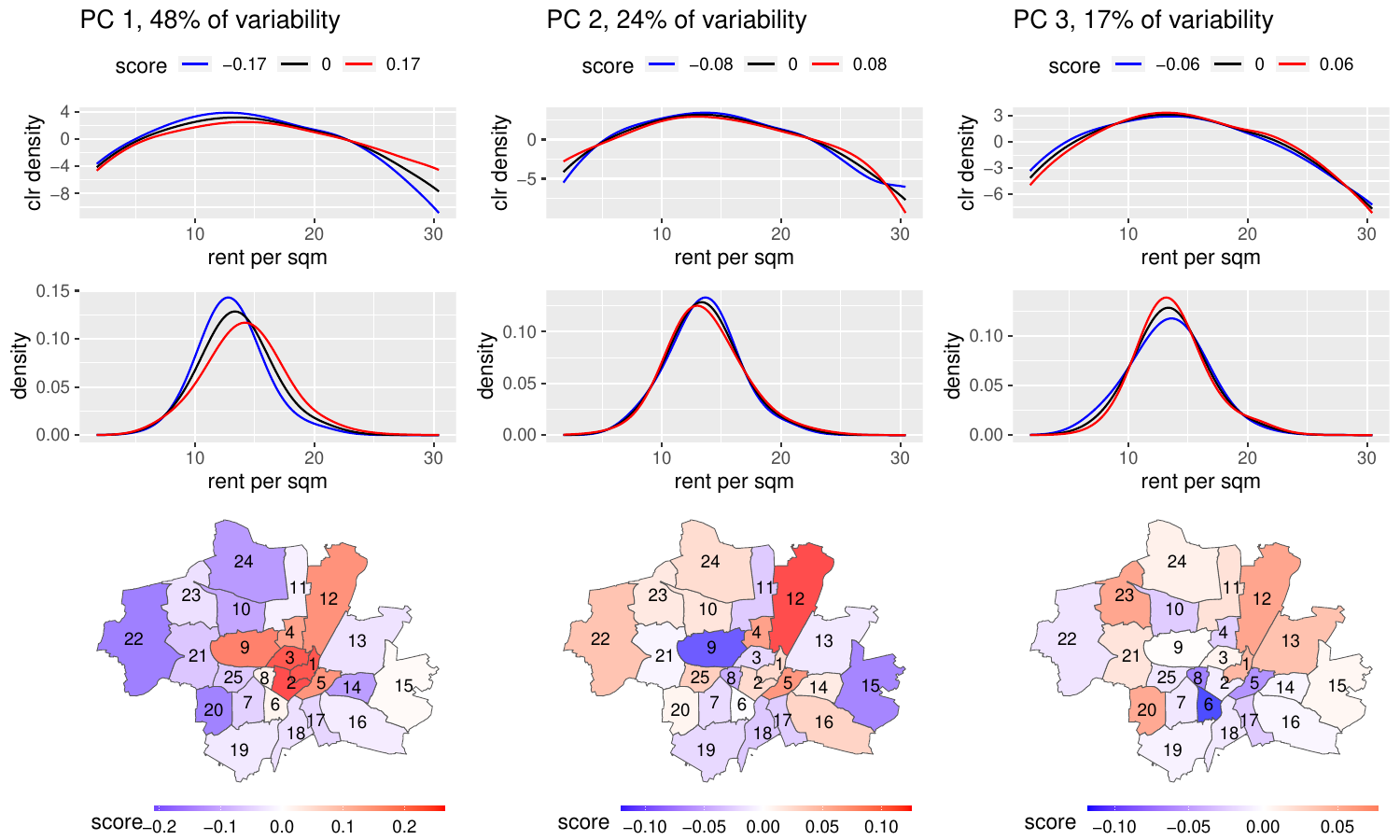}
   \caption{PCA based on the latent density model.}
   \label{fig:rent_denstiy_pca} 
\end{subfigure}\\[0.2cm]

\begin{subfigure}[c]{\textwidth}
   \includegraphics[width=\textwidth]{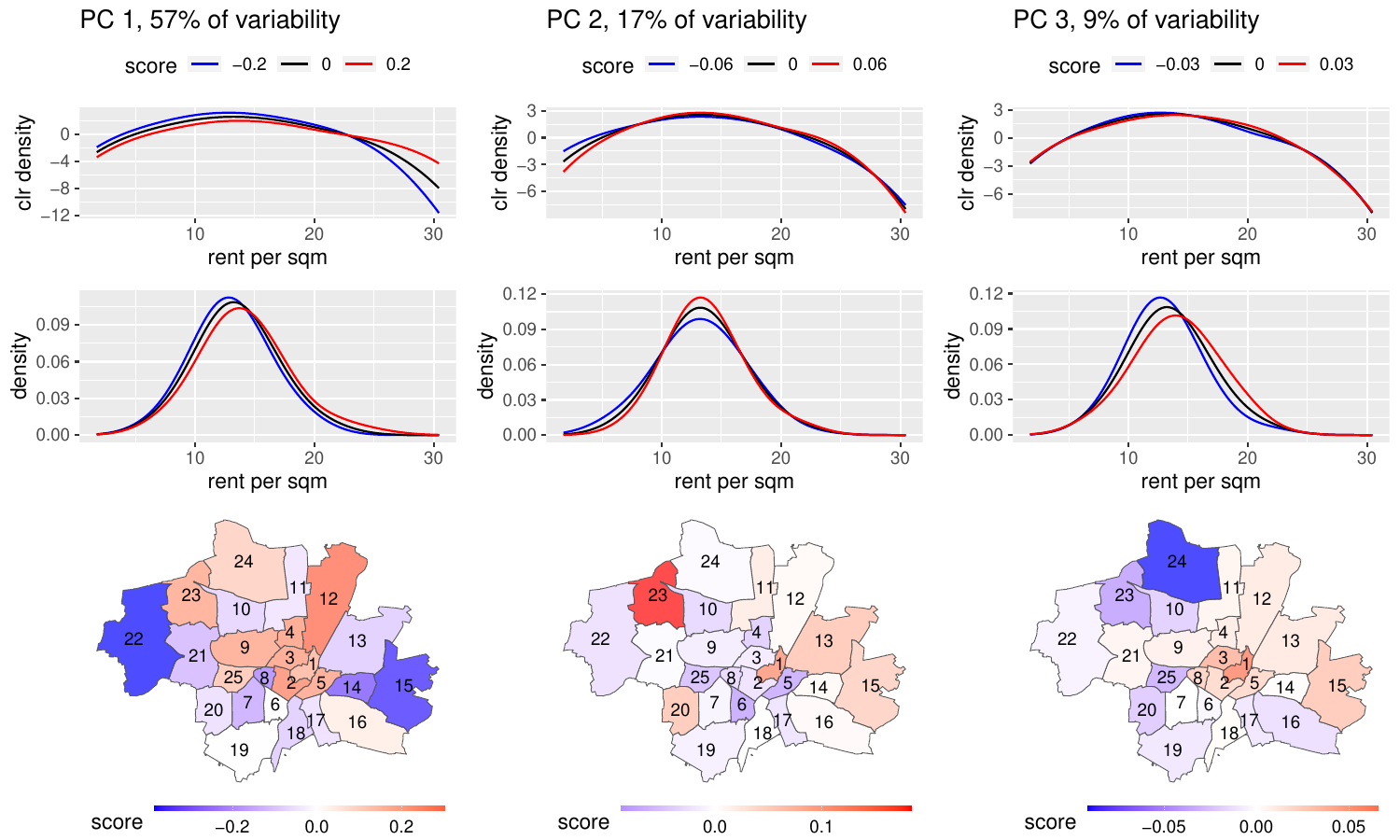}
    \caption{Two-step approach using kernel density estimates and then applying PCA after clr transformation.}
   \label{fig:rent_pca_pre_smooth}
\end{subfigure}
\caption[]{Comparison of the latent density model with a two-step approach using kernel density estimates as preprocessing on the Munich rent dataset. For each method, the first row shows the effect of adding a principal component times the corresponding standard deviation to the mean on clr level, the second row shows the same effect on density level, and the third row shows a map of Munich districts, where the color represents the predicted scores.}
\end{figure}

We first consider in Figure \ref{fig:rent_denstiy_pca} the results of estimating the latent density model with kernel density estimates as initial estimates. Details for the estimation can be found in Appendix \ref{app:rent}. In this model, the first principal component causes a shift towards more expensive apartments, and looking at the effect on clr level, we see that this principal component also primarily describes whether the occurrence of very expensive apartments (more than 25€ per square meter) is likely. Looking at which districts have a positive score for the first principal component, we see that these districts are mainly in the city center and correspond closely to the districts that have an increasing influence on the expected net rent in the regression model estimated in \cite{fahrmeir}.

The second principal component describes the presence of very cheap and expensive apartments. However, this principal component has only a small influence on the shape of the rent per square meter density. In the latent density model the variance of the principal components, and thus their importance, is measured at the clr level. Since in this case the largest effect is on parts of the clr transformed densities with negative values, the multiplicative effect on the actual densities is in parts where the density is close to zero and thus the effect, corresponding to heavier tails, is hardly visible. It is therefore not surprising that the spatial distribution of the corresponding scores has little structure, since the second principal component describes the occurrence of 
a few extreme observations in the districts.

The effect of the third principal component on the densities appears to be larger than the effect of the second principal component, although the variability explained is smaller, as it affects areas around the mode of the distribution. For this component, a negative score mostly describes a larger share of affordable housing (5€ to 10€ per square meter). These negative scores are mainly predicted for districts in the south of Munich, i.e.\ 6-Sendling and neighboring districts.

Comparing the effects of the first three principal components estimated with the latent density model with the estimates obtained using a two-step approach based on pre-smoothing with kernel density estimates (Gaussian kernel, bandwidth $= 2$), we see that the estimates differ considerably. In particular, for the two-step approach, the main variation in the scores is caused by density estimates of districts with only a small number of observations $m_i$. That is, the most extreme scores are estimated for district 22 with $m_{22} = 35$, district 23 with $m_{23} = 29$, and district 24 with $m_{24} = 38$, for the first three principal components, respectively. These districts benefit from the shrinkage effect of the latent density model, which divides the total variance into a part which is due to the underlying stochastic process for the latent densities and a part due to sampling from them. This shrinkage causes the predictions for the densities in these districts to be closer to the overall mean. Note also that the percentages of variance explained for the PCs shown in figures \ref{fig:rent_denstiy_pca} and \ref{fig:rent_pca_pre_smooth}, respectively, are correspondingly relative to different total variances.

Figure \ref{fig:rent_histograms} in the appendix shows that for these densities, the latent density predictions appear more plausible than the kernel density estimates, especially in areas where there are few observations, i.e. for very cheap or expensive housing. However, nearly all predicted densities appear to better reflect the underlying data using our approach, as the kernel density estimates generally underestimate the modes. The reason for this behavior is that we had to choose a relatively large bandwidth for the kernels to avoid estimating close to zero densities in other parts of the domain, especially for small samples. While we use the same kernel density estimates for our approach as starting values, the influence of such problems and the choice of the bandwidth in general is much smaller due to the later updates of the model.
In order to systematically investigate the influence of the number of observations per density on the model estimation, a simulation is carried out in the following subsection.
\section{Simulation} \label{sec:simulation}
This subsection aims to evaluate how well our latent density model can recover the mean and covariance structure of the latent process for a varying number of observations per density, ranging from very few observations to a moderate number. We also compare the performance of the model with two-step approaches, where the density estimates are obtained first, and then the PCA is performed after applying the clr transformation to each density. This corresponds to the simplicial PCA proposed by \cite{hron}. For the first comparison, we obtain kernel density estimates with a Gaussian kernel and then perform PCA on the clr transformed densities, which is also used as the initial estimate for our method. Second, we use a compositional spline  estimate for the clr transformed densities, as suggested by \cite{machalova}, before performing the PCA. 

\begin{figure}[ht]
    \centering
    \includegraphics[width=\textwidth]{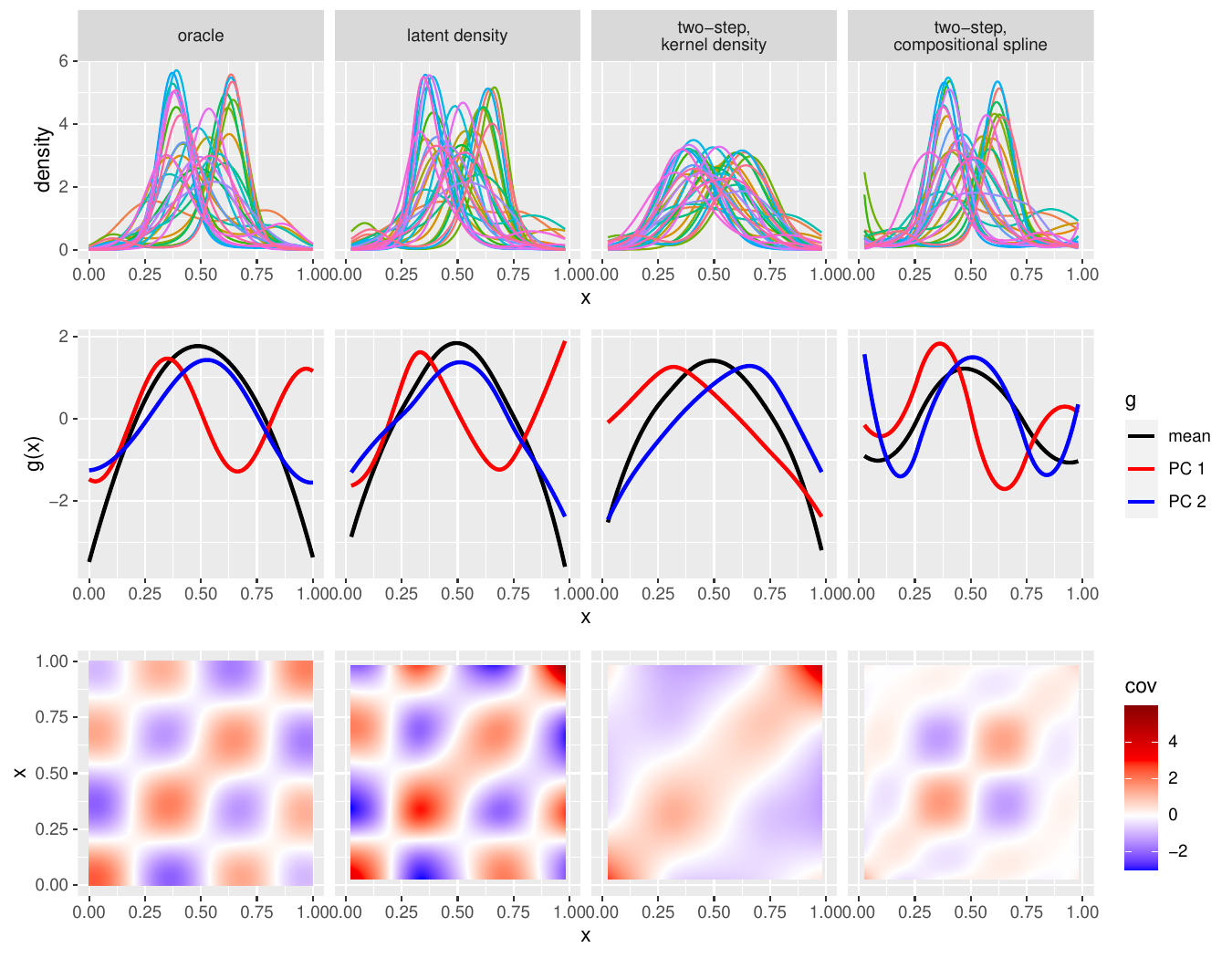}
    \caption{One randomly selected simulation run with $m_i = 40$ observations per density. Here 'oracle' refers to estimation based on the true underlying densities. Top: Density estimates/predictions. Middle: Mean and first two principal components. Bottom: Estimated covariance structure.}
    \label{fig:simulation_example}
\end{figure}

To this end we choose the following simulation setting. In each simulation run we simulate $n = 30$ densities on the interval $I = [0,1]$ from a Gaussian process with  true clr transformed mean function $\mu(x) = -20(x-\frac{1}{2})^2 + \frac{5}{3}$ and only two principal components. These are given on clr level as $g_1(x) = \frac{1}{5} \sin(10(x-\frac{1}{2}))$ with corresponding factor $Z_{1} \sim \mathcal{N}(0, 0.5)$ and $g_2(x) = \frac{1}{10} \cos(2\pi(x-\frac{1}{2}))$  with corresponding factor $Z_{2} \sim \mathcal{N}(0, 0.2)$. Note that these functions satisfy $\mu, g_1, g_2 \in \mathbb{L}^2_0$ and $g_1 \perp g_2$. The samples for the densities $f_i$, $i = 1, \dots, 30$ are then obtained as $f_i = \clr^{-1}(\mu + z_{i1} g_1 + z_{i2} g_2)$ where $z_{i1} \overset{i.i.d.}{\sim} Z_1$ and $z_{i2} \overset{i.i.d.}{\sim} Z_2$. The resulting densities are shown in Figure \ref{fig:simulation_example} in the top row on the left for a simulation run with $m_i = 40$. Finally, we sample observations $x_{ij} \overset{i.i.d}{\sim}f_i$ with $j = 1, \dots, m_i$ from each density $f_i$, $i = 1, \dots, n$. For the number of observations per density we consider $m_i \in \{20, 40, 80, 160\}$ and repeat the simulation  100 times for each $m_i$.

For the two-step approaches we then estimate the densities, either with kernel density estimates using Gaussian kernels with bandwidths $0.12, 0.09, 0.08, 0.07$ for the different setting with  $m_i = 20, 40, 80, 160$, respectively, or using cubic compositional splines with five knots. The kernel density estimates are also used as initial estimates for our latent density model. The number of Monte Carlo samples in the E step \ref{estep} is chosen to be $r = 10h$, where $h$ is the iteration index, i.e., the number of samples increases over the iterations. The parameter $\lambda$ for the proposal density is set to $1$. In each iteration, the dimension reduction is at most $0.0001$, i.e. we keep as many principal components as necessary to explain at least 99.999 \% of the variance.

\begin{figure}[ht]
    \centering
    \includegraphics[width=\textwidth]{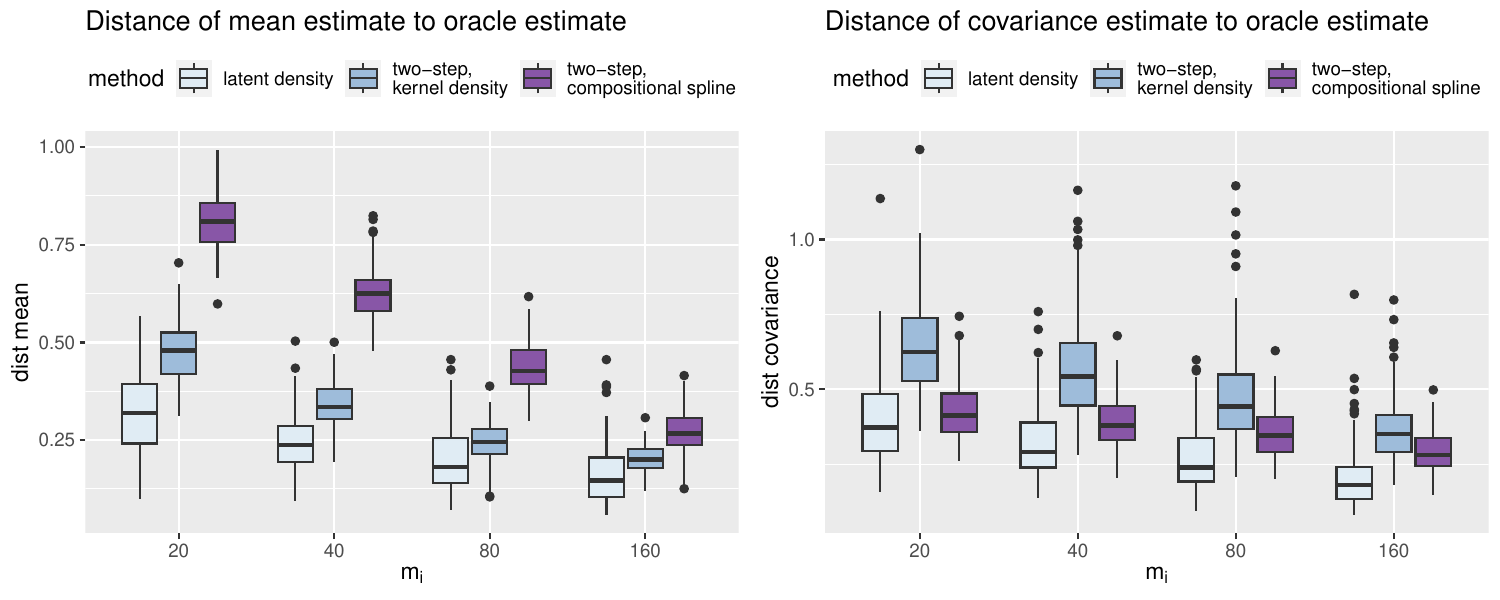}
    \caption{$\mathbb{L}^2$ distance of the estimated mean and covariance functions for each method to the oracle estimates based on the true underlying densities $f_1, \dots, f_n$.}
    \label{fig:simulation_dist_to_oracle}
\end{figure}

The performance of the different methods is evaluated as the distance to the oracle estimates, that are the pointwise estimates of the mean and the covariance function based on the true underlying densities $f_1, \dots, f_n$ (Figure \ref{fig:simulation_dist_to_oracle}) evaluated on a equidistant grid with 200 grid points. More precisely, we compute the distance of the mean estimates as $\sqrt{\int_0^1 (\tilde{\mu}(x) - \hat{\mu}(x))^2 dx}$, where $\tilde{\mu}$ is the oracle estimate for the mean and $\hat{\mu}$ is the estimate of each method. Analogously, the distance of the covariance functions is obtained as $\sqrt{\int_0^1 \int_0^1 (\tilde{C}(x_1, x_2) - \hat{C}(x_1, x_2))^2 dx_1 dx_2}$, where $\tilde{C}$ is the oracle estimate for the covariance and $\hat{C}$ is the estimate of each method.

As expected, the performance of all three methods improves as the number of observations per density increases. Still, over all values of $m_i \in \{20, 40, 80, 160\}$, and for both the mean and the covariance function, our latent density model has the smallest average distance to the oracle estimate over the 100 replicates. This shows that our method outperforms both two-step approaches in this scenario. However, when comparing the two-step approaches, it is worth noting that kernel density estimates seem to be better for estimating the mean, while compositional splines excel at capturing the covariance structure. For a more concrete picture of how the mean and covariance estimates for the three methods behave relative to the oracle estimate, Figure \ref{fig:simulation_example} shows each for an example with $m_i = 40$ observations per density. 

In this example, the estimates for the mean function, the first two principal components, and the covariance function of the latent density model (second column) appear similar to the corresponding oracle estimates (first column). Also, the pattern of predicted latent densities (first row, third column) is similar to the pattern of true underlying densities. For the two-step approach based on kernel density estimates (third row), the modes of the estimated densities appear to be too low, and although the mean has a similar shape as the oracle mean, the first two principal components, as well as the covariance function, differ substantially from the oracle estimates. 

The compositional splines used for the two-stage approach in the fourth row appear to behave similarly to the true densities near the center of the observed interval (in $\approx [0.2, 0.8]$), but show implausible characteristics near the boundary, i.e., near $0$ and $1$. This is also evident in the estimates for the mean and the first two principal components. Similarly, the estimate for the covariance function seems to be close to the oracle estimate in the center of the $I \times I$ domain, but not at the boundaries.

\section{Discussion} \label{sec:discussion}
We have proposed improvements to existing PCA methods for densities in the Bayes Hilbert space, by explicitly incorporating in a maximum likelihood approach that there are usually only discrete samples of the densities available.  This differs from two-step approaches, where the densities are first estimated in a preprocessing step and then PCA is performed for these estimates in the Bayes Hilbert space, ignoring uncertainty arising from the first density estimation step. We confirmed in applications and in a simulation that our latent density model can be successfully estimated employing an MCEM algorithm, and that the estimation of the mean and covariance structure for the densities is superior to the estimation using two-step approaches. While improvements are particularly pronounced for small samples per density, differences persist even for moderately large samples.

Consequently, resulting estimates for the principal components using our approach are better suited for understanding the variation in the underlying densities, and the predicted scores can be better used for dimension reduction and for further analyses, such as to describe differences and trends in the densities. In addition, given the importance of principal components for dimension reduction in functional data analysis \citep{ramsay_silverman, chiou}, the principal components could also be used as the basis for further subsequent functional data analysis methods, such as as a basis in which to expand model terms in a functional regression model \citep[as for example in][]{yao, scheipl, volkmann} in the Bayes Hilbert space. Finally, the predicted latent densities could also be used for subsequent analysis, providing more reliable reconstructions of the underlying true densities than the density estimates obtained using  preprocessing with kernel density estimation or compositional splines.

To allow maximum likelihood estimation of the principal components, we assume a normal distribution of the scores, i.e. a Gaussian process (prior) for the latent densities, as common in many statistical methods ranging from mixed models \citep{guo} to Gaussian process regression \citep{rasmussen}. This distributional assumption may be too restrictive in some applications, however, in such cases leading to unjustified shrinkage of the latent densities. In future research, it would thus be appealing to extend our approach further by including a choice of different distributions for the scores in the latent density model, e.g. to account for heavy tails.

In this paper, our primary focus was on continuous densities with respect to the Lebesgue measure on a bounded interval. Furthermore, we provided a brief outline of a potential extension to discrete measures, which incorporates compositional data.  In addition to an application of our model to the compositional data case often encountered in practical data scenarios, an interesting direction for future research would be a generalization of the approach and implementation to other measures. This could include mixed discrete and continuous measures as in \cite{maier}, the inclusion of unbounded domains as well as the multivariate case.

\section*{Acknowledgements}
We gratefully acknowledge funding by grants GR 3793/3-1 `Flexible regression methods for curve and shape data' and GR 3793/8-1 `Flexible density regression methods' from the German Research Foundation (DFG). We would also like to thank the data providers in the ECA\&D project for providing the temperature data for the first application and Michael Windmann for sharing the Munich rent data for the second application.

\clearpage

\bibliographystyle{plainnat}
\bibliography{literature}

\newpage
\appendix
\section{Proofs and Computations}
\subsection{Proof of Lemma \ref{lem:clr_trafo}} \label{app:clr_trafo}
\begin{proof}
The clr transformation is well defined as 
\begin{align*}
    \left| \int_I \log(f(x)) dx \right| \leq |I| \left\| \log(f) \right \|_{\mathbb{L}^2} < \infty,
\end{align*}
due to the Cauchy-Schwarz inequality and $\log(f) \in \mathbb{L}^2$. We further compute 
\begin{align*}
    \clr([\alpha f]) &= \log(\alpha f) - \frac{1}{|I|} \int_I \log(\alpha f(x)) dx \\
    &= \log(\alpha) + \log(f) - \frac{1}{|I|} \int_I \log(\alpha) + \log(f(x)) dx \\
    &= \log(f) - \frac{1}{|I|} \int_I \log(f(x)) dx = \clr([f])
\end{align*}
for all $\alpha \in \mathbb{R}$ and $f \in B$. Since $\clr^{-1}$ is clearly well defined as well, we show that $\clr$ is bijective via showing
\begin{align*}
    \clr(\clr^{-1}(g)) = \clr([\exp(g)]) = g - \frac{1}{|I|} \int_I g(x) dx = g
\end{align*}
for all $g \in \mathbb{L}_0^2$, 
and
\begin{align*}
    \clr^{-1}(\clr(f)) = \left[\exp\left(\log(f) - \frac{1}{|I|} \log(f(x)) dx \right)\right] = \left[f \exp\left( - \frac{1}{|I|} \log(f(x)) dx \right)\right] = [f] 
\end{align*}
for all $f \in B$. The clr transformation is also an isometry as for all $g_1, g_2 \in \mathbb{L}_0^2$ holds
\begin{align*}
    \langle [\clr^{-1}(g_1)], [\clr^{-1}(g_2)] \rangle_\mathcal{B} =& \frac{1}{2|I|} \int_{I} \int_{I} \log\left( \frac{\exp(g_1(x))}{\exp(g_1(y))} \right) \log\left( \frac{\exp(g_2(x))}{\exp(g_2(y))} \right) dx \ dy \\
    =& \frac{1}{2|I|} \int_{I} \int_{I} \left(g_1(x) - g_1(y) \right) \left( g_2(x) - g_2(y) \right) dx \ dy \\
    =& \frac{1}{2|I|} \int_{I} \int_{I} \left( g_1(x)g_2(x) - g_1(y)g_2(x) - g_1(x)g_2(y) + g_1(y)g_2(y) \right) dx \ dy \\
    =& \frac{1}{2} \int_{I} g_1(x)g_2(x) dx - \frac{1}{2|I|} \int_{I} g_1(y) dy \int_I g_2(x) dx - \frac{1}{2|I|} \int_{I} g_1(x) dx \int_I g_2(y) dy \\
    & +\frac{1}{2} \int_{I} g_1(y)g_2(y) dy \\
    =& \int_{I} g_1(x)g_2(x) dx = \langle g_1, g_2 \rangle_{\mathbb{L}^2}
\end{align*}
since $\int_{I} g_1(x) dx = \int_I g_2(x) dx = 0$.
\end{proof}
\vfill

\subsection{Proof of Corollary \ref{cor:finite_gauss_process}} \label{app:finite_gauss_process}
\begin{proof}
It is trivial to see that if $G = \sum_{k = 1}^N \theta_k e_k$ with $\boldsymbol{\theta} \overset{i.i.d.}{\sim} \mathcal{N}(\boldsymbol{\nu}, \boldsymbol{\Sigma})$,
then $G \overset{i.i.d.}{\sim} GP(\mu, K)$ with $\mu = \sum_{k = 1}^N \nu_k e_k$ and $K(x_1, x_2) = \sum_{k = 1}^N \sum_{l = 1}^N e_k(x_1) e_l(x_2) \boldsymbol{\Sigma}_{kl}$ since 

\begin{align} \label{eq:covariance}
\mathbb{E}(G(x)) &=  \mathbb{E}\left( \sum_{k = 1}^N \theta_k e_k(x)\right) =
\sum_{k = 1}^N \nu_k e_k \quad \text{and} \nonumber \\
\text{Cov}(G(x_1), G(x_2)) &= \text{Cov}\left( \sum_{k = 1}^N \theta_k e_k(x_1),  \sum_{l = 1}^N \theta_l e_l(x_2)\right) =
\sum_{k = 1}^N \sum_{l = 1}^N  e_k(x_1)  e_l(x_2) \text{Cov}(\theta_k , \theta_l) \nonumber \\
&= \sum_{k = 1}^N \sum_{l = 1}^N e_k(x_1) e_l(x_2) \boldsymbol{\Sigma}_{kl}.
\end{align}

The other direction is an implication of the Karhunen-Loève decomposition. If $\mathcal{H}$ is $N'$-dimensional, $G \overset{i.i.d.}{\sim} GP(\mu, K)$ can be decomposed as $G = \mu + \sum_{k = 1}^{N'} Z_k \varphi_k$ with $\varphi_k$, $k = 1, \dots, {N'}$ being the orthonormal eigenfunctions and uncorrelated scores $Z_k$ with $\mathbb{E}(Z_k) = 0$ and Var$(Z_k) = \sigma_k^2$.

Choose  $\nu_k = \langle \mu, e_k \rangle_{\mathbb{L}_2}$ to be the orthonormal projection on $e_k$ and $\boldsymbol{\Sigma}_{kl} = \sum_{j = 1}^{N'} \sigma_j^2 \langle \varphi_j, e_k \rangle_{\mathbb{L}_2}\langle \varphi_j, e_l \rangle_{\mathbb{L}_2}$ for all $k,l = 1, \dots, N$. Then we compute
\begin{align*}
    \sum_{k = 1}^N \nu_k e_k &= \sum_{k = 1}^N \langle \mu, e_k \rangle_{\mathbb{L}_2} e_k = \mu, 
\end{align*}
since this gives the orthogonal projection of $\mu$ on span$\{e_1, \dots, e_N\}  $ and $\mu \in \mathcal{H} \subseteq \text{span}\{e_1, \dots, e_N\}$ . We further compute using the same identity
\begin{align*}
    \text{Cov}(G(x_1), G(x_2)) &= \text{Cov}\left( \sum_{k = 1}^{N'} Z_k \varphi_k(x_1),  \sum_{l = 1}^{N'} Z_l \varphi_l(x_2)\right) =
    \sum_{k = 1}^{N'} \sum_{l = 1}^{N'}  \varphi_k(x_1)  \varphi_l(x_2) \text{Cov}(Z_k , Z_l) \\
    &= \sum_{k = 1}^{N'} \sigma_k^2 \varphi_k(x_1)  \varphi_k(x_2)
\end{align*}
which is identical to
\begin{align*}
\sum_{k = 1}^N \sum_{l = 1}^N  e_k(x_1)  e_l(x_2) \boldsymbol{\Sigma}_{kl} 
&= \sum_{k = 1}^N e_k(x_1) \sum_{l = 1}^N e_l(x_2) \sum_{j = 1}^{N'}  \sigma_j^2 \langle \varphi_j, e_k \rangle_{\mathbb{L}_2}\langle \varphi_j, e_l \rangle_{\mathbb{L}_2} \\
&= \sum_{j = 1}^{N'}  \sigma_j^2 \sum_{k = 1}^N e_k(x_1)  \langle \varphi_j, e_k \rangle_{\mathbb{L}_2} \sum_{l = 1}^N e_l(x_2) \langle \varphi_j, e_l \rangle_{\mathbb{L}_2} \\
&= \sum_{j= 1}^{N'}  \sigma_j^2 \varphi_j(x_1)  \varphi_j(x_2).
\end{align*}
To show the correspondence of the eigenvalue decompositions we need to show that if $\boldsymbol{v}_l = (v_{l1}, \dots, v_{lN})$ is an eigenvector of $\boldsymbol{\Sigma}$ with corresponding eigenvalue $\sigma^2_l$ then  $\varphi_l = \sum_{m = 1}^N v_{lm} e_m$ is an eigenfunction of $K$ with the same eigenvalue $\sigma^2_l$. This is true since if we plug in the formula  obtained in \eqref{eq:covariance} for $K$ we obtain 

\begin{align*}
    \int_I K(x_1, \cdot) \varphi_l(x_1) dx_1 
    &= \int_I K(x_1, \cdot) \sum_{m = 1}^N v_{lm} e_m(x_1) dx_1 =
    \sum_{k = 1}^N e_k \int_I \left( \sum_{j = 1}^N e_j(x_1) \boldsymbol{\Sigma}_{kj} \right) \sum_{m = 1}^N v_{lm} e_m(x_1) dx_1 \\
    &= \sum_{k = 1}^N e_k \sum_{j = 1}^N   \boldsymbol{\Sigma}_{kj} \sum_{m = 1}^N v_{lm} \int_I e_j(x_1) e_m(x_1) dx_1
    = \sum_{k = 1}^N e_k \sum_{j = 1}^N   \boldsymbol{\Sigma}_{kj} v_{lj} \\
    &= \sum_{k = 1}^N e_k \left(\boldsymbol{\Sigma} \boldsymbol{v}_l  \right)_k 
    = \sum_{k = 1}^N e_k \left(\sigma^2_l \boldsymbol{v}_l  \right)_k
    = \sigma^2_l \sum_{k = 1}^N e_k v_{lk} = \sigma^2_l \varphi_l.
\end{align*} 

\subsection{Derivation of the likelihood} \label{app:likelihood}
\begin{align*}
   L(\mu, K| \boldsymbol{x}_i, \dots, \boldsymbol{x}_n) &= \prod_{i = 1}^n  p(\boldsymbol{x}_i| \mu, K) = \prod_{i = 1}^n \int_{\mathbb{R}^N} p(\boldsymbol{x}_i| \boldsymbol{\theta}_i) p(\boldsymbol{\theta}_i | \mu, K) d \boldsymbol{\theta}_i \\
   &= \prod_{i = 1}^n \int_{\mathbb{R}^N} \left( \prod_{j = 1}^{m_i} p(x_{ij}| \boldsymbol{\theta}_i) \right) p(\boldsymbol{\theta}_i | \mu, K) d \boldsymbol{\theta}_i \\
   &= \prod_{i = 1}^n  \int_{\mathbb{R}^N}  \left( \prod_{j = 1}^{m_i} \frac{\exp(\sum_{k = 1}^N \theta_{ik} e_k(x_{ij}))}{\int_I \exp(\sum_{k = 1}^N \theta_{ik} e_k(x)) dx}  \right) p(\boldsymbol{\theta}_i | \mu, K) d \boldsymbol{\theta}_i \\
   &= \prod_{i = 1}^n  \int_{\mathbb{R}^N} \frac{ \exp( \sum_{j = 1}^{m_i} \sum_{k = 1}^N \theta_{ik} e_k(x_{ij})) p(\boldsymbol{\theta}_i | \mu, K) }{\left(\int_I \exp(\sum_{k = 1}^N \theta_{ik} e_k(x)) dx\right)^{m_i}}  d \boldsymbol{\theta}_i
\end{align*}
\end{proof}

\subsection{Proof of Lemma \ref{lem:proper_density}} \label{app:proper_density}
\begin{proof}
To show this statement we consider the three non constant additive parts of the logarithm of the conditional distribution $\log(p(\mathbf{z}_i | \boldsymbol{x}_i, \boldsymbol{\nu}^{(h)}, \boldsymbol{\Sigma}^{(h)}))$. 
\begin{itemize}
    \item The first part $\sum_{j = 1}^{m_i} \left( \mu^{(h)}(x_{ij}) + \sum_{k = 1}^N \boldsymbol{z}_i^T \boldsymbol{v}^{(h)}_k  e_k(x_{ij}) \right)$ is linear in $\boldsymbol{z}_i$, which means there is a constant $M_1 \in \mathbb{R}$ such that it is $M_1 \boldsymbol{z}_i$.
    \item The second part is always negative, as
    \begin{align*}
        -m_i\log\left( \int_I \exp\left( \mu^{(h)}(x) + \sum_{k = 1}^N \boldsymbol{z}_i^T \boldsymbol{v}^{(h)}_k  e_k(x)  \right) dx \right) \leq -m_i \int_I \mu^{(h)}(x) + \sum_{k = 1}^N \boldsymbol{z}_i^T \boldsymbol{v}^{(h)}_k  e_k(x) dx = 0
    \end{align*}
    where the inequality is due to Jensen's inequality and the integral is equal to zero as all functions are in $\mathbb{L}^2_0$.
    \item The third part consists of normal densities and is therefore quadratic in $\boldsymbol{z}_i$. More precisely, we have
    \begin{align*}
        \log(\prod_{k = 1}^N p(z_{ik}| {\sigma_k^2}^{(h)}))
        = \sum_{k = 1}^N \frac{-z_{ik}^2}{2 {\sigma_k^2}^{(h)}} + \text{const.} \leq - \frac{1}{{2 \sigma_1^2}^{(h)}} \|\boldsymbol{z}_i\|^2 + \text{const.}
    \end{align*}
\end{itemize}
Taking these three parts together this shows that there are constants $M_1, M_2 \in \mathbb{R}$ and $M_3 > 0$ such that $\log(p(\mathbf{z}_i | \boldsymbol{x}_i, \boldsymbol{\nu}^{(h)}, \boldsymbol{\Sigma}^{(h)})) \leq - M_3  \|\boldsymbol{z}_i\|^2 + M_1 \boldsymbol{z}_i + M_2$, which implies that $p(\mathbf{z}_i | \boldsymbol{x}_i, \boldsymbol{\nu}^{(h)}, \boldsymbol{\Sigma}^{(h)})$ is a proper density.
\end{proof}

\subsection{Example: The posterior mode will not necessarily be attained if the prior is improper} \label{app:counterex_no_mode}
\begin{proof}
Consider the $1$-dimensional case $N = 1$ with densities defined on the unit interval $I = [0,1]$ and only one basis function for the clr transformed densities given as $e_1 = \mathds{1}_{[0,0.5]} - \mathds{1}_{]0.5,1]}$. This means also the parameter space $\mathbb{R}$ in this case and assuming for this parameter $z_i \in \mathbb{R}$ an improper $1$-dimensional normal distribution is equivalent to assuming a flat prior.

If we further assume there is only one observation $x_{i1} = 0.2$ and for the prior mean holds $\boldsymbol{\nu}^{(h)} = 0$ we compute
\begin{align*}
    p(z_i | x_{i1}, \boldsymbol{\nu}^{(h)}, \boldsymbol{\Sigma}^{(h)}) &\propto \frac{\exp(z_i e_1(x_{i1}))}{\int_{[0,1]} \exp(z_i e_1(x)) dx} = \frac{\exp(z_i)}{\int_{[0,0.5]} \exp(z_i) dx + \int_{[0.5, 1]} \exp(-z_i) dx} \\
    &= \frac{2 \exp(z_i)}{\exp(z_i) + \exp(-z_i)} = \frac{2}{1 + \exp(-2 z_i)},
\end{align*}
which is monotonously increasing in $z_i \in \mathbb{R}$. Hence it does not attain its maximum and therefore $p(z_i | x_{i1}, \boldsymbol{\nu}^{(h)}, \boldsymbol{\Sigma}^{(h)})$ does not define a proper distribution in this case.
\end{proof}

\subsection{Proof of Lemma \ref{lem:grad}} \label{app:grad}
Like in the proof of Lemma \ref{lem:proper_density} (Appendix \ref{app:proper_density}) we consider the three non constant additive parts of the logarithm of the conditional distribution $\log(p(\mathbf{z}_i | \boldsymbol{x}_i, \boldsymbol{\nu}^{(h)}, \boldsymbol{\Sigma}^{(h)}))$. 
\begin{itemize}
    \item For the linear part $\sum_{j = 1}^{m_i} \left( \mu^{(h)}(x_{ij}) + \sum_{k = 1}^N \boldsymbol{z}_i^T \boldsymbol{v}^{(h)}_k  e_k(x_{ij}) \right)$ the gradient with respect to $\boldsymbol{z}_i$ is given as
    \begin{align*}
        \frac{\partial}{\partial \boldsymbol{z}_i} \sum_{j = 1}^{m_i} \sum_{k = 1}^N \boldsymbol{z}_i^T \boldsymbol{v}^{(h)}_k  e_k(x_{ij}) =  \sum_{j = 1}^{m_i} \sum_{k = 1}^N \boldsymbol{v}^{(h)}_k  e_k(x_{ij}) =  \sum_{k = 1}^N \boldsymbol{v}^{(h)}_k \sum_{j = 1}^{m_i}  e_k(x_{ij})
    \end{align*}
    
    \item For the second part we compute the gradient as
    \begin{align*}
        &\phantom{=} \frac{\partial}{\partial \boldsymbol{z}_i} -m_i\log\left( \int_I \exp\left( \mu^{(h)}(x) + \sum_{k = 1}^N \boldsymbol{z}_i^T \boldsymbol{v}^{(h)}_k  e_k(x) \right) dx \right) \\
        &=  \frac{-m_i \int_I \frac{\partial}{\partial \boldsymbol{z}_i} \exp\left( \mu^{(h)}(x) + \sum_{k = 1}^N \boldsymbol{z}_i^T \boldsymbol{v}^{(h)}_k  e_k(x) \right) dx }{\int_I \exp\left( \mu^{(h)}(x) + \sum_{k = 1}^N \boldsymbol{z}_i^T \boldsymbol{v}^{(h)}_k  e_k(x) \right) dx} \\
        &= \frac{-m_i \int_I \exp\left( \mu^{(h)}(x) + \sum_{k = 1}^N \boldsymbol{z}_i^T \boldsymbol{v}^{(h)}_k  e_k(x) \right)  \sum_{k = 1}^N \boldsymbol{v}^{(h)}_k  e_k(x) dx }{\int_I \exp\left( \mu^{(h)}(x) + \sum_{k = 1}^N \boldsymbol{z}_i^T \boldsymbol{v}^{(h)}_k  e_k(x) \right)dx } \\
        &= -m_i \int_I \frac{\exp\left( \mu^{(h)}(x) + \sum_{k = 1}^N \boldsymbol{z}_i^T \boldsymbol{v}^{(h)}_k  e_k(x) \right)}{\int_I \exp\left( \mu^{(h)}(x) + \sum_{k = 1}^N \boldsymbol{z}_i^T \boldsymbol{v}^{(h)}_k  e_k(x) \right)dx }  \sum_{k = 1}^N \boldsymbol{v}^{(h)}_k  e_k(x) dx \\
        &= -m_i \int_I \clr^{-1}\left( \mu^{(h)} + \sum_{k = 1}^N \boldsymbol{z}_i^T \boldsymbol{v}^{(h)}_k  e_k \right)(x) \sum_{k = 1}^N \boldsymbol{v}^{(h)}_k  e_k(x) dx \\
        &= -m_i \sum_{k = 1}^N \boldsymbol{v}^{(h)}_k \int_I f_{\mathbf{z}_i}(x)  e_k(x) dx
    \end{align*}
     where, in the first equation, we can interchange differentiation and integration applying the Leibniz rule, since we have assumed that all clr transformed densities are bounded.
    \item The third part is the sum of logarithms of normal densities. Therefore, for all $l = 1, \dots, N$ we compute the partial derivative with respect to $z_{il}$ as
    \begin{align*}
        \frac{\partial}{\partial z_{il}}\log(\prod_{k = 1}^N p(z_{ik}| {\sigma_k^2}^{(h)}))
        = \frac{\partial}{\partial z_{il}} \frac{-z_{il}^2}{2 {\sigma_l^2}^{(h)}} = \frac{-z_{il}}{{\sigma_l^2}^{(h)}}.
    \end{align*}
\end{itemize}
Adding these three parts together gives the gradient of the logarithm of the conditional density of the scores.

\section{Additional plots for the applications in Section \ref{sec:applications}} \label{app:additional_plots}

\subsection{Temperature data} \label{app:temperature}
For the temperature data, the latent density model is estimated using kernel density estimates with a Gaussian kernel and bandwidth $=1.5$ as initial estimates. The number of Monte Carlo samples in the E step \ref{estep} is chosen to be $r = 50h$, where $h$ is the iteration index, i.e., the number of samples increases over the iterations. The tuning parameter $\lambda$ for the proposal density is set to the default value $1$. In each iteration, the dimension reduction is at most $0.001$, i.e. we keep as many principal components as necessary to explain at least 99.99 \% of the variance.

\begin{figure}[ht]
    \centering
    \includegraphics[width=\textwidth]{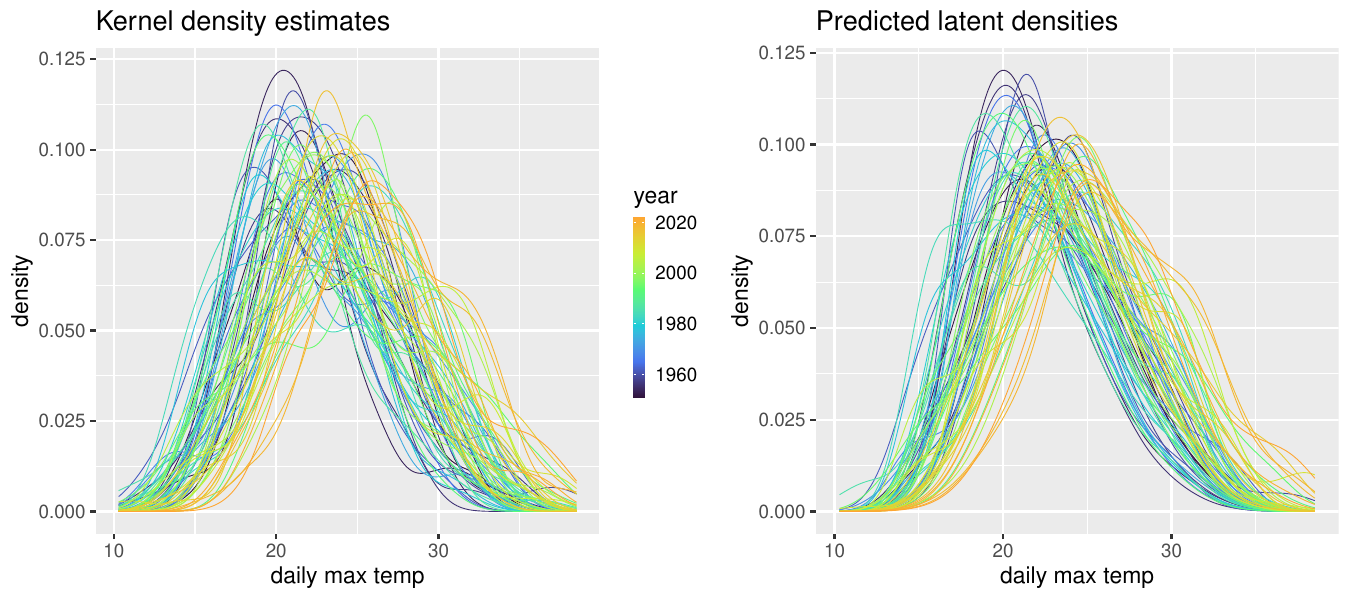}
    \caption{Kernel density estimates (Gaussian kernel, bandwidth $= 1.5$) and predicted latent densities estimated from the latent density model \eqref{eq:latent_density_model} for the daily maximum summer temperatures in Berlin Tempelhof from 1951 to 2022. The kernel densities are also used as initial estimates in the latent density model.}
    \label{fig:temperature_densities}
\end{figure}

\begin{figure}[ht]
    \centering
    \includegraphics[width=\textwidth]{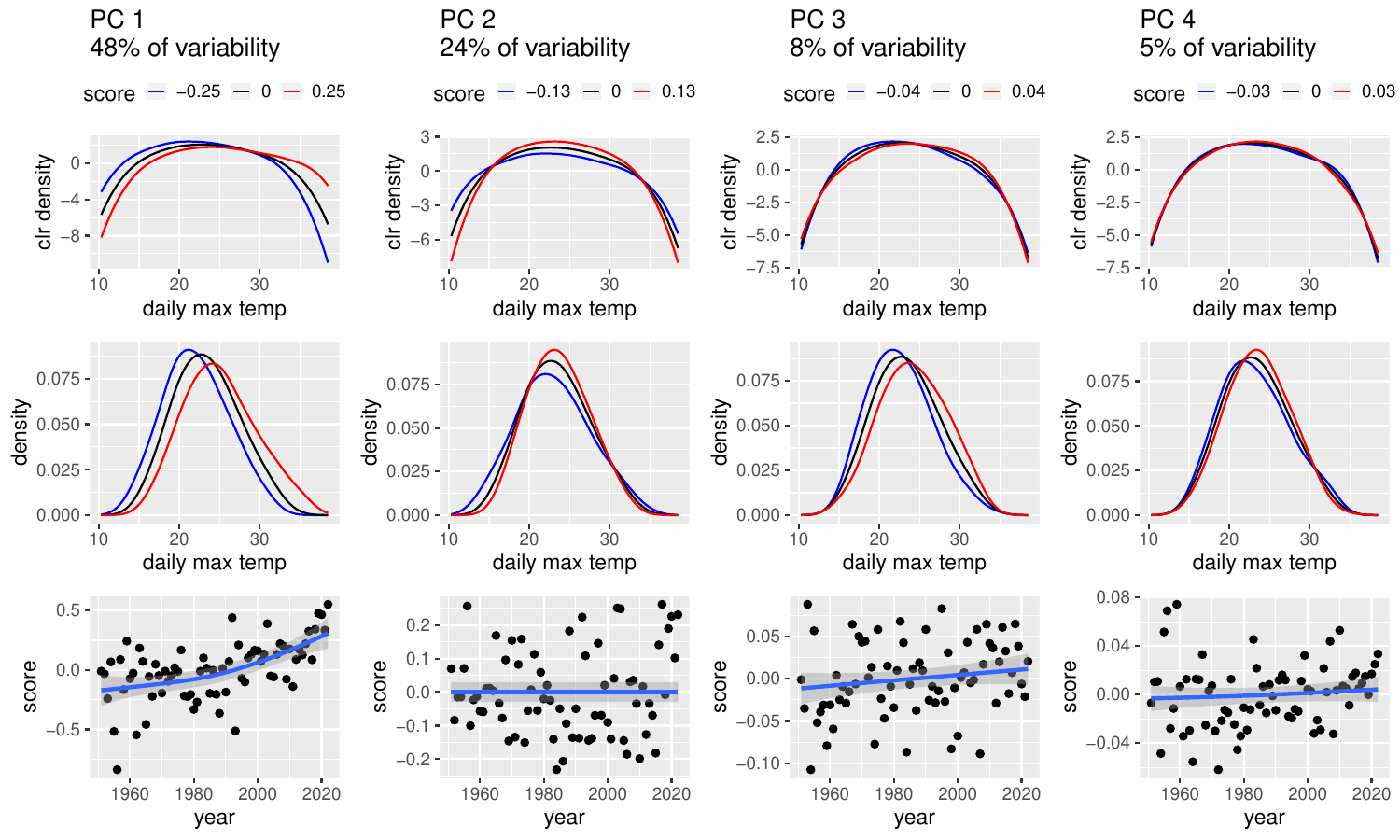}
    \caption{PCA based on the clr transformations of the kernel density estimates (Gaussian kernel, bandwidth $= 1.5$) for daily maximum temperature (°C). Top: Effect of adding/subtracting $\hat{\sigma}_k \hat{\varphi}_k$ to the clr transformed mean density $\mu$, where $\hat{\varphi}_k$ is the $k$th principal component, with corresponding eigenvalue $\hat{\sigma}_k^2$, $k = 1,2,3, 4$. Middle: Effect on the density level, i.e. $\clr^{-1}$ transformations of the functions in the top row. Bottom: Temporal trend of the corresponding predicted scores per year, with scatterplot smoother and pointwise confidence bands overlaid.}
    \label{fig:temperature_pca_pre_smooth}
\end{figure}

In Figure \ref{fig:temperature_densities} we show the kernel density estimates and the predicted latent densities obtained from the latent density model. For both, the trend towards higher temperatures is evident. Note that the variance of the predicted latent densities is smaller than the variance of the kernel density estimates, since the latent density model effectively splits the total variance into the variance due to the underlying stochastic process for the latent densities and the variance due to sampling from them. Correspondingly, percentages variance explained for the PCs depicted in Figures \ref{fig:temperature_density_pca} and \ref{fig:temperature_pca_pre_smooth}, respectively, are relative to different total variances.
\vfill
\newpage

\vfill
\newpage
\subsection{Rental prices} \label{app:rent}

For the rent index data, the latent density model is estimated using kernel density estimates with a Gaussian kernel and bandwidth $= 2$ as initial estimates. The number of Monte Carlo samples in the E step \ref{estep} is chosen to be $r = 100h$, where $h$ is the iteration index, i.e., the number of samples increases over the iterations. The parameter $\lambda$ for the proposal density is set to $2$. In each iteration, the dimension reduction is at most $0.0005$, i.e. we keep as many principal components as necessary to explain at least 99.995 \% of the variance.

\begin{figure}[ht]
    \centering
    \includegraphics[width=\textwidth]{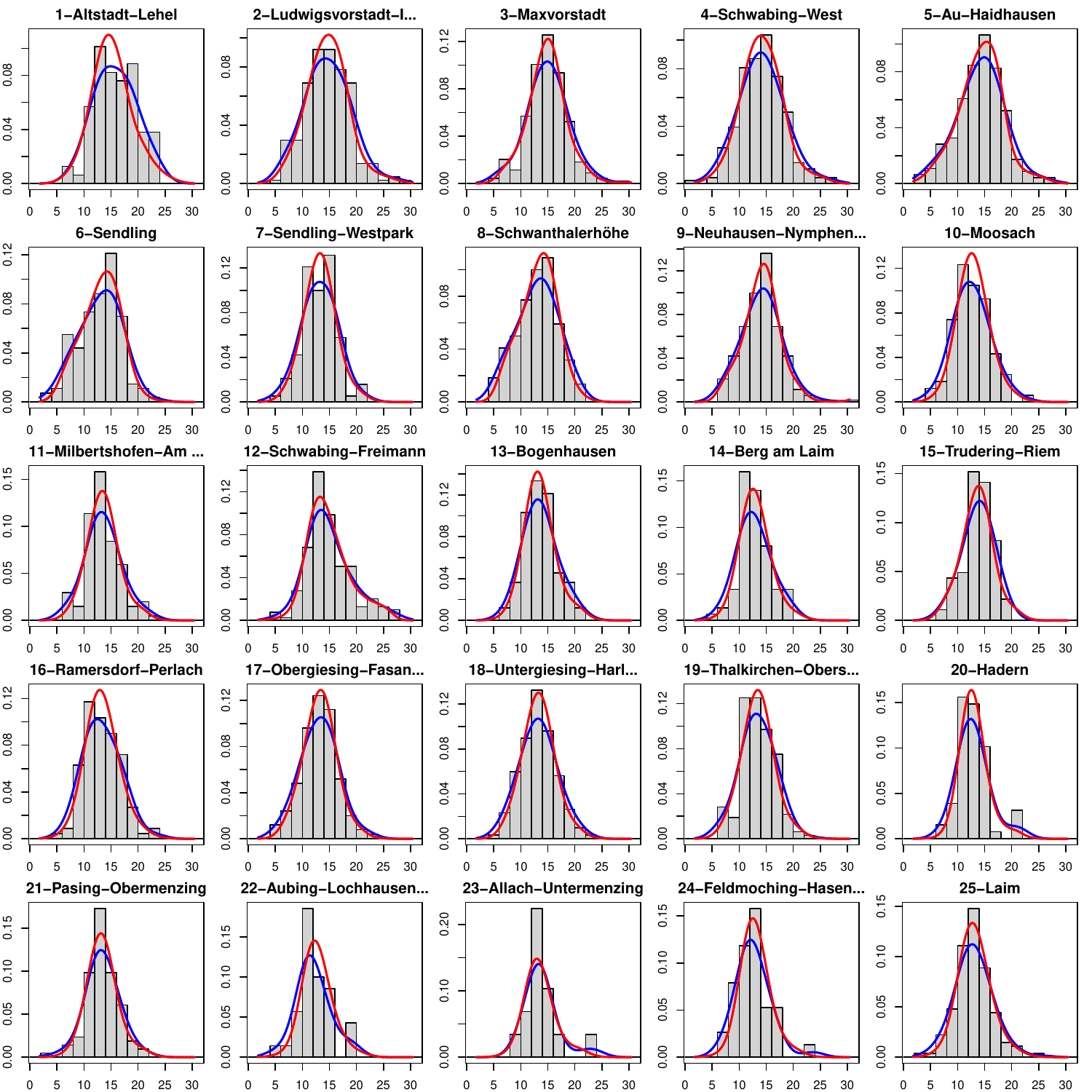}
    \caption{Histograms of rent per square meter for each district, with overlaid kernel density estimates (Gaussian kernel, bandwidth $= 2$) in blue and predicted latent densities based on our proposed approach in red.}
    \label{fig:rent_histograms}
\end{figure}

\end{document}